\documentclass{emulateapj}
\usepackage{graphicx}
\input psfig.sty
\usepackage{multirow}
\usepackage{textcomp}
\usepackage{epsfig}
\usepackage{amsmath}
\usepackage{amssymb}
\usepackage{amsthm}
\usepackage{xy}
\shorttitle{A timing and spectral model for the AXPs XTE J1810 and CXOU J1647}
\shortauthors{Albano et al.}

\begin{document}

\title{An unified timing and spectral model for the Anomalous X-ray Pulsars\\
XTE J1810-197 and CXOU J164710.2-455216}

\author{A. Albano\altaffilmark{1,2}, R. Turolla\altaffilmark{1,3},
G.L. Israel\altaffilmark{2}, S. Zane\altaffilmark{3}, L.
Nobili\altaffilmark{1} and L. Stella\altaffilmark{2}}
\altaffiltext{1}{Department of Physics,
University of Padova, Via Marzolo 8, I-35131 Padova, Italy}
\altaffiltext{2}{INAF-Astronomical Observatory of Rome, via
Frascati 33, I-00040, Monte Porzio Catone, Italy}
\altaffiltext{3}{Mullard Space Science Laboratory, University
College London, Holmbury St. Mary, Dorking, Surrey, RH5 6NT, UK}

\begin{abstract}

Anomalous X-ray pulsars (AXPs) and soft gamma repeaters (SGRs) are
two small classes of X-ray sources strongly suspected to host a magnetar, i.e.
an ultra-magnetized neutron star with $B\approx 10^{14}$--$10^{15}$ G.
Many SGRs/AXPs are known to be variable, and recently the existence of
genuinely ``transient'' magnetars was discovered.
Here we present a comprehensive study of the pulse profile and spectral evolution
of the two transient AXPs (TAXPs) XTE J1810-197 and CXOU J164710.2-455216.
Our analysis was carried out in the framework of the twisted magnetosphere model for magnetar
emission. Starting from 3D Monte Carlo simulations of the emerging
spectrum, we produced a large database of synthetic pulse profiles
which was fitted to observed lightcurves in different
spectral bands and at different epochs. This allowed us to derive
the physical parameters of the model and their evolution with time,
together with the geometry of the two sources, i.e. the
inclination of the line-of-sight and of the magnetic axis with
respect to the rotation axis. We then fitted the (phase-averaged)
spectra of the two TAXPs at different epochs using a model similar
to that used to calculate the pulse profiles ({\tt ntzang} in XSPEC)
freezing all parameters to the values obtained from the timing
analysis, and leaving only the normalization free to vary. This
provided acceptable fits to {\it XMM-Newton} data in all the
observations we analyzed. Our results support a picture in which a
limited portion of the star surface close to one of the magnetic
poles is heated at the outburst onset. The subsequent evolution is
driven both by the cooling/varying size of the heated cap and by a
progressive untwisting of the magnetosphere.

\end{abstract}


\section{Introduction}\label{intro}

In recent years an increasing number of high-resolution spectral and timing
observations of isolated neutron stars has become available.
Many of these observations concern two peculiar classes of high-energy
pulsars, the Anomalous X-Ray Pulsars (AXPs: 9 objects plus 1 candidate)
and the Soft Gamma-Ray Repeaters (SGRs: 6 objects)
\footnote{see {\tt http://www.physics.mcgill.ca/\textasciitilde pulsar/magnetar/main.html}
for an updated catalogue of SGRs/AXPs}.
Historically these two classes of sources were regarded as distinct.
While SGRs were first discovered in late 1978-early 1979, when SGR 1806-20 and SGR 0526-66
exhibited a bright burst of soft $\gamma$-rays
\citep{mazets79, laros86}, AXPs were observed for the first time in 1981,
when \cite{fg81} discovered pulsations in the {\it EINSTEIN} source 1E 2259+586.
It was, however, not until the mid '90s that AXPs were recognized as a class
of ``anomalous'' pulsars because of their luminosity
substantially exceeding rotational energy losses \citep{merste95}.

Although SGRs were mainly known as emitters of short,
energetic bursts, they are also persistent X-ray sources with
properties quite similar to those of AXPs \cite[see e.g.][for
reviews]{woodsthomp06,mereghetti08}. They all are slow X-ray
pulsars, with spin periods in a very narrow range ($P \sim 2$--12
s), relatively large spin-down rates ($\dot{P} \sim
10^{-13}$--$10^{-11} \ {\rm s \ s}^{-1}$), spin down ages of
$10^{3}$--1$0^{4} \ {\rm yr}$, and stronger magnetic fields compared
to those of rotation or accretion powered pulsars ($B \sim 10^{14}$--$10^{15} \
{\rm G} > B_{QED} \simeq 4.4 \times 10^{13} \ {\rm G}$). AXPs and
SGRs have persistent X-ray luminosities $L_X\sim
10^{34}$--$10^{36} \ {\rm erg \ s}^{-1}$.
%, considerably higher than those of
%other isolated neutron stars presently known.
Their spectra in the 0.1--10 keV band are relatively soft and can be empirically
fitted with a two-component model, an absorbed blackbody
($kT \sim 0.2$--0.6 keV) plus a power-law ($\Gamma \sim 2$--4).
{\it INTEGRAL} observations revealed the presence of
sizeable emission up to $\sim 200$ keV, which accounts for up to
$50\%$ of the total flux. Hard X-ray spectra are well represented
by a power-law, which dominates above $\sim 20$ keV in AXPs.

The large high-energy output can not be explained in terms of rotational
energy losses, as in conventional models for radio-pulsars, while
the lack of stellar companions argues against accretion.
The powering mechanism of AXPs and SGRs, instead, is believed to
reside in the neutron star ultra-strong magnetic field (magnetar;
\citealt{duncan92}, \citealt{thompson93}).
The magnetar scenario appears capable to explain the properties
of both the bursts \citep{td95} and the persistent emission
\cite[the twisted magnetosphere model,][and references therein; see
\S~\ref{model} for details]{thompson02, zrtn09}, although
no definite model for the hard tails was put forward as yet.

The persistent emission of SGRs and AXPs is now known to be variable.
AXPs, in particular, display different types of X-ray flux variability:
from slow, moderate flux changes on timescales of months/years,
to intense outbursts with short rise times ($\sim 1$ day) lasting
$\sim 1$ year. Some AXPs were found to undergo
intense and dramatic SGR-like burst activity on sub-second
timescales (XTE J1810-197, 4U 0142+614, 1E 1048.1-5973, CXOU
J164710.2-455216 and 1E 2259+586). The discovery of bursts from
AXPs is regarded as further evidence in favor of a common nature
of AXPs and SGRs.

The first case of AXP flux variability was observed in 2002, when
1E 2259+586 showed an increase in the persistent flux by a factor
$\sim 10$ with respect to the quiescent level, followed by the
emission of $\sim  80$ short bursts with luminosity $L_{X} \sim
10^{36}-10^{38} \ {\rm erg \ s}^{-1}$ \citep{kaspi03}.
In early 2003 the 5.54 s
AXP XTE J1810-197 was discovered at a luminosity $\sim 100$
greater than its quiescent value \cite[$10^{33} \ {\rm erg \ s}^{-1}$;][]{ibrahim04}.
Analysis of archival data revealed that
the outburst started between November 2002 and January 2003.

On September 21st 2006 an outburst was observed from the AXP CXOU
J164710.2-455216 ($P = 10.61$ s). The flux level was $\sim 300$
times higher than that measured only 5 days earlier by {\it
XMM-Newton} \citep{muno06b, campana06, israel06}. This, much as in
the case of XTE J1810-197, indicates that some AXPs are transient
sources (dubbed Transient AXPs) and may become
visible only when they enter an active state. Recently other AXPs
and SGRs showed a series of short bursts of soft $\gamma$-rays
which was detected by different satellites
\citep{mereghetti09}.

In this paper we present a comprehensive study of the pulse profile
and spectral evolution of the TAXPs XTE J1810-197 and CXOU J164710.2-455216
throughout their outbursts of November 2002 and September 2006, respectively.
By confronting timing data with synthetic lightcurves obtained from
the twisted magnetosphere model \citep{nobili08}, we were able to estimate
how the physical parameters of the source (surface temperature and
emitting area, electron energy, twist angle) evolve in time.
The fits of the pulse profiles also allowed us to infer the geometry
of the two systems, i.e. the angles between the magnetic and rotational
axes and the line of sight. Spectral models, obtained with the
parameter values derived from the timing analysis, provide
acceptable fits to {\it XMM-Newton} data.

\section{Transient AXPs properties}\label{axps}

\subsection{XTE J1810-107}\label{xte}

The Transient AXP (TAXP) XTE J1810-197 was serendipitously
discovered in 2003 with the Rossi X-Ray Timing Explorer ({\it
RXTE}) while observing SGR 1806-20
(\citealt{ibrahim04}). The source was readily identified as a
X-ray pulsar, and soon after a search in archival {\it RXTE} data
showed that it produced an outburst around 2002 November, followed
by a monotonic decline of the X-ray flux. The X-ray pulsar spin
period was found to be 5.54 s, with a spin-down rate $\sim
10^{-11} \ {\rm s \ s}^{-1}$. Using the standard expression for
magneto-rotational losses, the inferred value of the (dipolar)
magnetic field is $B\sim 3 \times 10^{14} \ {\rm G}$
\citep{ibrahim04}. The source was classified as the first
transient magnetar. The TAXP XTE J1810-197 was then studied with
{\it Chandra} and {\it XMM-Newton}
\cite[][]{gotthelf04,israel04,gotthelf05,gotthelf07}, in
order to monitor its evolution in the post-outburst phase.

By using archival Very Large Array ({\it VLA}) data a transient
radio emission with a flux of $\sim 4.5 \ {\rm mJy}$ at $1.4 \
{\rm GHz}$ was discovered at the {\it Chandra} X-ray position of
XTE J1810-197 \citep{halpern05}. Only later on it was
discovered that this radio emission was pulsed, highly polarized
and with large flux variability even on  very short timescales
(\citealt{camilo06}). The X-ray and the radio pulsations are at
the same rotational phase. Since accretion is expected to quench
radio emission, this is further evidence against the source being
accretion-powered.

Deep IR observations were performed for this source, revealing
a weak ($K_{s} = 20.8$ mag) counterpart, with characteristics similar to
those of other AXPs \citep{israel04}. The IR emission is variable
\citep{rea04b}, but no correlations between the IR and X-ray changes
were found up to now. The existence of a correlation at IR/radio
wavelengths is uncertain \citep{camilo06,camilo07a, testa08}.

XTE J1810-197 was observed 9 times by {\it XMM-Newton}, between
September 2003 and September 2007, two times every year. The
uninterrupted coverage of the source during 4 years provides an
unique opportunity to understand the phenomenology of TAXPs.
Earlier observations of XTE J1810-197 showed that the source spectrum
is well reproduced by a two blackbody model, likely indicating
that (thermal) emission occurs in two regions of the star
surface of different size and temperature: a hot one ($kT = 0.70 \
{\rm keV}$) and a warm one ($kT = 0.30 \ {\rm keV}$;
\citealt{gotthelf05}). {\it XMM-Newton} observations also showed
that the pulsed fraction decreases in time.

\cite{perna08} discussed the post-outburst spectral evolution
of XTE J1810-197 from 2003 to 2005 in terms of two blackbody components,
one arising from a hot spot and the other from a warm concentric ring.
By varying the area and temperature of the two regions, this (geometric) model
can reproduce the observed spectra, account for the decline of
the pulsed fraction with time and place a strong
constrain on the geometry of the source, i.e. the angles
between the line of sight and the hot spot axis with respect to
the spin axis.

Recently, \cite{bernardini09} by re-examining all
available {\it XMM-Newton} data found that inclusion of a third
spectral component, a blackbody at $\sim 0.15$ {\rm keV}, improved the fits.
When this component is added both the area and temperature of the hot
component was found  to monotonically decrease in time,
while the warm component decreased in area but stayed at constant temperature.
The coolest blackbody, which appeared not to change in time, is associated
to emission from the (large) part of the surface which was not affected
by the event which triggered the outburst, and is
consistent with the spectral properties of the source as derived
from a {\it ROSAT} detection before the outburst onset. Finally,
an interpretation of XTE J1810-197 spectra in terms of a resonant
compton scattering model (RCS, see \S \ref{model}) was presented by
\cite{rea08}.

\subsection{CXOU J164710.2-455216}\label{cxou}

The TAXP CXOU J164710.2-455216 was discovered in two {\it Chandra}
pointings of the young Galactic star cluster Westerlund 1 in
May/June 2005. The period of the X-ray pulsar was found to be
$P=10.61$ s \citep{muno06}, with a period derivative $\dot{P}=9
\times 10^{-13} \ {\rm s \ s}^{-1}$ \citep{israel07}. The implied
magnetic field is $B \sim 10^{14} \ {\rm G}$.

In November 2006, an intense burst was detected by the {\it Swift}
Burst Alert Telescope (BAT) in Westerlund 1
\citep{krimm06, muno06}. Its short duration ($20 \
{\rm ms}$) suggested that its origin was the candidate AXP.
However the event was initially attributed to a nearby Galactic
source, so the AXP was not promptly re-observed
by {\it Swift}. A ToO observation program with {\it Swift} was
started 13 hrs after the burst, displaying a persistent flux level $300$ times
higher than the quiescent one. CXOU J164710.2-455216 was observed
in radio, with the Parkes Telescope. The observation was carried
out a week after the outburst onset, with the intent of searching
for pulsed emission similar to that of XTE J1810. In this
case, however, only a (tight) upper limit to the radio flux ($40 \
\mu {\rm Jy}$) was placed \citep{burgay06}.

{\it XMM-Newton} observations carried out across the outburst onset show
a complex pulse profile evolution. Just before the event the pulsed
fraction was $\sim 65\%$, while soon after it became
$\sim 11 \%$ \citep{muno07}. Moreover, the pulse profile changed
from being single-peaked just before the burst, to showing three peaks
soon after it. CXOU J164710.2-455216 spectra in the outburst state were
fitted either with a two blackbody model ($kT_1\sim 0.7 \ {\rm keV}$,
$kT_2\sim 1.7 \ {\rm keV}$), or with  a blackbody plus power-law model
\cite[$kT \sim 0.65 \ {\rm keV}$, $\Gamma \sim 2.3$; ][]{muno07,israel07}.
\cite{rea08} found that a RCS model also provides a good fit to the data.

\section{The model}\label{model}

It is now widely accepted that AXPs and SGRs are magnetars, and
that their burst/outburst activity, together with the persistent
emission, are powered by their huge magnetic field. In particular,
the soft X-ray spectrum ($\sim 1$--10 keV) is
believed to originate in a ``twisted'' magnetosphere
\cite[$\nabla\times{\mathbf B}\neq 0$;][]{thompson02}, where the
currents needed to support the field provide a large enough
optical depth to resonant Compton scattering of thermal photons
emitted by the star surface. Since charges are expected to flow along
the closed field lines at relativistic velocities, photons gain energy
in the (resonant) scatterings and ultimately fill a hard tail.

Most studies on spectral formation in
a twisted magnetosphere \citep{lg06,ft07,nobili08} are based on
the axially symmetric, force-free solution for a twisted dipolar
field presented by \cite{thompson02}. This corresponds to a
sequence of magnetostatic equilibria which, once the polar
strength of the magnetic field $B_p$ is fixed, depends only on a
single parameter: the radial index of the magnetic field $p$
($B\propto r^{-p-2}$, $0\le p\le 1$) or, equivalently, the twist
angle
\begin{equation}
 \Delta\phi_{N-S} = \lim_{\theta_{0} \to 0}2\int_{\theta_{0}}^{\pi/2}
\frac{B_{\phi}}{B_{\theta}}\frac{d\theta}{\sin\theta}\,,
\end{equation}
where $B_r, B_\theta$ and $B_\phi$ are the spherical components of the field,
which depend only on $r$ and $\theta$ because of axial symmetry.
Knowledge of ${\mathbf B}$ fixes the current density ${\mathbf j}=
(c/4\pi)\nabla\times{\mathbf B}$, and, if the particle
velocity is known, also the electron density in the magnetosphere
\begin{equation}
 n_{e}(r,\theta) = \frac{p+1}{4\pi e}\Bigl(\frac{B_{\phi}}{B_{\theta}}\Bigr)
\frac{B}{r |\langle \beta \rangle|}
\end{equation}
where $e$ is the electron charge and $\langle \beta \rangle$ is the
average charge velocity (in units of $c$).
The charge density of the space charge-limited flow of ions and electrons
moving along the closed field lines is orders of magnitude larger
than the Goldreich-Julian density, $n_{GJ}$, associated to the charge
flow along the open field lines in radio-pulsars.

In our investigation we make use of the spectral models presented
by \citet[NTZ in the folowing]{nobili08}, who studied radiative transfer in a
globally twisted magnetosphere by means of a 3D Monte Carlo code.
Each model is characterized by the magnetospheric twist
$\Delta\phi_{N-S}$, the electron (constant) bulk velocity $\beta$,
and the seed photon temperature $kT$. The polar field was fixed at
$B_p=10^{14} \ {\rm G}$. In the applications presented by
NTZ, it was assumed that the star surface emits
unpolarized, blackbody radiation and is at uniform temperature.
Concerning the present investigation, the most critical assumption
is that of a globally twisted magnetosphere, as discussed in some
more detail in \S~\ref{discus}. Taking a constant value for the
electrons bulk velocity is certainly an oversimplification and
reflects the lack of a detailed model for the magnetospheric
currents. In a realistic case one would expect that $\beta$ is a
function of position. However, resonant scattering is possible
only where the bulk velocity is mildly relativistic. If along a
flux tube there are large variations of the Lorentz factor, only
the region where $\beta\approx 0.5$ will contribute to scattering.
Moreover, preliminary calculations of the dynamics of charged
particles in a twisted force-free magnetosphere performed
accounting for both electrostatic acceleration and Compton drag
indicate that $\beta$ is indeed fairly constant along the central
 part of a flux tube (Beloborodov, private communication).
The assumption of unpolarized
thermal radiation is not cogent either, since we are not
interested in the polarization of the escaping radiation and the
emergent spectrum is quite insensitive to the polarization
fraction of the seed photons (see, e.g., Fig. 4 of NTZ).

\begin{figure}
\includegraphics[width=3.3in,angle=0]{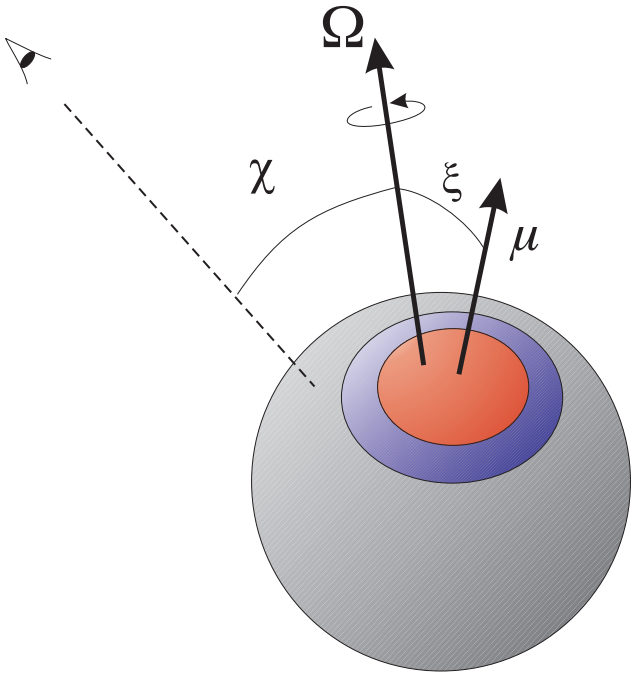}
\caption{A schematic view of the neutron star.
${\mathbf\Omega}$ and
${\mathbf\mu}$ are the star spin and magnetic axis, respectively.
The dashed line corresponds to the line-of-sight. The two angles
$\chi$ and $\xi$ are also shown. The star surface is divided
into three regions: a hot polar cap (red), a warm corona (blue)
and a colder zone (gray). \label{mag1}}
\end{figure}

The code works
by dividing the stellar surface into $N_{\Theta} \times N_{\Phi}$
zones of equal area by means of a ($\cos \Theta$, $\Phi$) grid,
where $\Theta$ is the magnetic colatitude and $\Phi$ the
longitude. After a few scatterings photons escape from the neutron
star magnetosphere and are collected on a spherical surface (the
``sky'') which is divided into $N_{\Theta_s} \times N_{\Phi_s}$
patches, similarly to what is done for the star surface. The key
point is that the evolution of seed photons from each patch is
followed separately. This allows us to treat an arbitrary surface
temperature distribution without the need to perform new Monte
Carlo runs, by simply combining together models from runs with
different temperatures at the post-production level
(the geometry is shown in Fig. \ref{mag1})

Monte Carlo models are computed (and stored) for the simplest
geometrical case, in which the spin and the magnetic axes are
aligned. As discussed in NTZ, the most general situation in which
the spin and magnetic axes are at an arbitrary angle $\xi$ can be
treated at the post-production level. If $\chi$  is the
inclination of the line-of-sight (LOS) with respect to the star
spin axis and $\alpha$ is the rotational phase angle, the
co-ordinates of the points where the LOS intersects the sky can be
found in terms of $\xi$, $\chi$ and $\alpha$. The pulse profile in
any given energy band is then obtained by integrating over the
selected range the energy-dependent counts at these
positions as the star rotates (see again NTZ for details). In
order to compare model lightcurves with observations, integration
over energy is performed by accounting for both interstellar
absorption and the detector response function. Actually, the
interstellar absorption cross-section $\sigma$ and the response
function $A$ depend on the photon energy at infinity $\bar
E=E\sqrt{1-R_S/R_{NS}}$, where $E$ is the energy in the star frame
(which is used in the Monte Carlo calculation) and $R_S$ is the
Schwarzschild radius (we assume a Schwarzschild space-time and take
$R_{NS} = 10 \ {\rm km}$ and $M_{NS}= 1.44 \ M_{\odot}$). Our model pulse profile in the $[\bar E_1,\, \bar
E_2]$ energy band is then proportional to
\begin{equation}\label{pulsep}
\int_{\bar E_1}^{\bar E_2}d\bar E\, \exp{[-N_H\sigma(\bar
E)]}A(\bar E) N(\alpha, E)
\end{equation}
where $N_H$ is the hydrogen column density and $N(\alpha, E)$ is
the phase- and energy-dependent count rate. In the applications
below we used the \cite{mormc83} model for interstellar absorption
and, since we deal with {\it XMM-Newton} observations, we adopted
the EPIC-pn response function. We remark that the Monte Carlo
spectral calculation is carried out assuming a flat space-time
(i.e. photons propagate along straight lines), so that, apart from
the gravitational redshift, no allowance is made for
general-relativistic effects \cite[see][for a more detailed
discussion]{zt06}. In particular, no constraints on the star mass
and radius can be derived in the present case from the comparison
of model and observed pulse profiles \cite[see e.g.][]{le08,le09}.

Finally, phase-averaged spectra are computed by summing over
all phases the energy-dependent counts.
Note that $0\le\xi\le\pi/2$, while $\chi$ is in the range $[0, \pi]$
because of the asymmetry between the north and south magnetic poles
introduced by the current flow.

\section{TAXP Analysis}\label{tanal}

Our first step in the study of the two TAXPs XTE J1810-197 and
CXOU J164710.2-455216 was to reproduce the pulse profiles (and
their time evolution) within the RCS model discussed in
\S\ref{model}. The fit to the observed pulse profiles in different
energy bands (total: $0.5 \ {\rm keV} \le  E \le 10 \ {\rm keV}$, soft:
$0.5 \ {\rm keV} \le  E \le 2 \ {\rm keV}$,
hard: $2 \ {\rm keV} \le  E \le 10 \ {\rm keV}$)
provides an estimate of the source parameters,
including the two geometrical angles $\xi$ and $\chi$. While the
twist angle, electron velocity and surface temperature may vary in
the different observations (although they must be the same in the
different energy bands for a given observation), the fits have to
produce values of $\xi$ and $\chi$ which are at all epochs
compatible with one another (to within the errors) in order
to be satisfactory. We then
computed the phase-averaged spectra for the two sources at the
various epochs for the same sets of parameters and compared them
with the observed ones. There are several reasons which led us to
choose such an approach. The main one is that, as discussed in NTZ
\cite[see also][]{zrtn09}, spectral fitting alone is unable to
constrain the two geometrical angles. Moreover, lightcurve fitting
allows for a better control in the case in which the surface
thermal map is complex and changes in time (see below).

For the present investigation, a model archive was generated
beforehand. Each model was computed by evolving $N_{patch} =
225,000$ photons for $N_{\Theta} \times N_{\Phi} = 8 \times 4 =
32$ surface patches ($N_{tot} = 7,200,000$ photons). The parameter
grids are: $0.1 \le kT \le 0.9 \ {\rm keV}$ (step $0.05 \ {\rm
keV}$), $0.1 \le \beta \le 0.9$ (step $0.1$) and $0.2\, {\rm rad} \le
\Delta\phi_{N-S} \le\, 1.2\, {\rm rad}$ (step $0.1\, {\rm rad}$). Photons are
collected on a $N_{\Theta_{s}} \times N_{\Phi_{s}} = 10 \times 10
= 100$ angular grid on the sky, and in $N_E = 50$ energy bins,
equally spaced in $\log E$ in the range $0.1-100 \ {\rm keV}$.

The analysis proceeds as follows. We first used the principal
component analysis (PCA) to explore the properties of the
lightcurves as a population and to select the model within the
archive that is closest to the observed one at a given epoch. This
serves as the starting point for the pulse profile fitting
procedure, which we performed by assuming that the whole star surface
is at the same temperature. The fitting is then repeated first for
the case in which the surface thermal distribution consists of a
hot spot and a cooler region, and then by generating a new archive
with a finer surface gridding, and applying it in the case of a
surface thermal map consisting of a hot spot, a warm corona and a
cooler region (see again Fig \ref{mag1}). Finally, the source parameters derived from the
lightcurve fitting are used to confront the model and observed
(phase-averaged) spectra. Phase-resolved spectral analysis, although feasible in our model
and potentially important, was not attempted
because the decay in flux of both sources makes the counting statistics rather poor after
the first one/two observations
\cite[see][for more details in the case of XTE J1810-197]{bernardini09}.

\subsection{PCA}\label{pca}

The principal component analysis is a method of multivariate
statistics that allows to reduce the number of variables $X_i$
needed to describe a data set by introducing a new set variables,
the principal components (PCs) $Z_i$. The PCs are linear
combinations of the original variables and are such that $Z_1$
displays the largest variance, $Z_2$ the second largest, and so
on. By using the PCs it is possible to describe the data set in terms of a
limited number of variables, which however, carry most of the
information contained in the original sample \citep[see e.g.][and references therein]{zt06}.

\begin{figure}
\includegraphics[width=3.3in,angle=0]{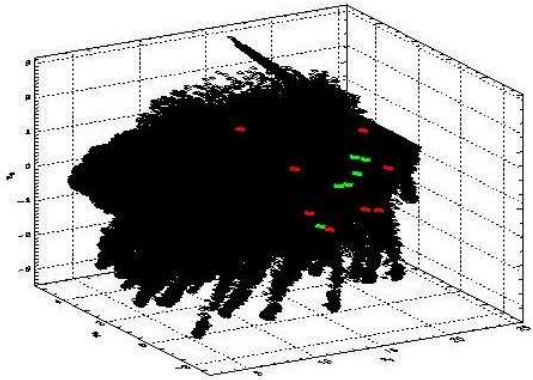}
\caption{Principal component representation  of the simulated
lightcurves in our archive (black squares) together with the observed
lightcurves of XTE J1810-197 (red dots) and of CXOU
J164710.2-455216 (green dots). All the pulse profiles refer to the
0.5--10 keV band. \label{pca_all_archive}}
\end{figure}

Synthetic lightcurves were generated for 32 phases in the
range $[0, 2\pi]$ and for a $9\times 9$ angular grid, $0^{\circ}
\le\xi\le 90^{\circ}$ (step $10^{\circ}$), $0^{\circ}\le\chi\le 180^{\circ}$
(step $20^{\circ}$); the archive contains a total of 136323 models.
Once the PCA was applied to the lightcurve set, we found that
the first three PCs ($Z_1, Z_2, Z_3$) accounts for as much as
$\sim 90 \%$ of the sample variance. This means that the entire
set is satisfactorily described in terms of just three variables
instead of the original 32 \cite[see][for an interpretation of
$Z_1,Z_2,Z_3$]{zt06}. A graphic representation of the
lightcurves in the archive in terms of the first three PCs is
shown in Fig. \ref{pca_all_archive}. In the same plot we also show
the PC representation of the pulse profiles of XTE J1810-197 and
CXOU J164710.2-455216 at the various epochs. The points
corresponding to observations fall within the volume occupied by
models and this guarantees that there is a combination of the
parameters for which a synthetic pulse profile reproduces the
data.  The PC representation is also used to find the model in the
archive which is closest to a given observed lightcurve, by looking
for the minimum of the (squared) Euclidean distance $\sum_{i=1}^{32}(Z_i-Z_i^{obs})^2$
between the model and the observed pulse profile.

\subsection{XTE J1810-197}\label{1810anal}

We considered eight {\it XMM-Newton} observations, covering the period
September 2003-September 2007 for the TAXP prototype XTE J1810-197
(see Table \ref{XTEobs} for the observation log). Only EPIC-pn data were used, and we refer
to \cite{bernardini09}, who analyzed the same observations, for all details on data extraction and
reduction. All the EPIC-pn spectra were rebinned before fitting, to have at least 40 counts per bin and
prevent oversampling the energy resolution by more than a factor of three.

\begin{deluxetable*}{lccccc}
\tablecolumns{6} \tablewidth{0pc} \tablecaption{XTE J1810-197
{\it XMM-Newton} observations\tablenotemark{a} \label{XTEobs}}
%\tablenum{1}
\tablehead{ \colhead{Label} & \colhead{OBS ID} & \colhead{Epoch} &
\colhead{Exposure time (s)} & \colhead{total counts} & \colhead{background counts} }
%\\\&  &  & & \colhead{tot ph} & \colhead{bck ph}}
\startdata
Sep03 & 0161360301 & 2003-09-08 & 5199 & 60136 & 2903 \\
Sep04 & 0164560601 & 2004-09-18 & 21306 & 89082 & 1574 \\
Mar05 & 0301270501 & 2005-03-18 & 24988 & 54279 & 1760 \\
Sep05 & 0301270401 & 2005-09-20 & 19787 & 21876 & 1311 \\
Mar06 & 0301270301 & 2006-03-12 & 15506 & 12296 & 1197 \\
Sep06 & 0406800601 & 2006-09-24 & 38505 & 23842 & 2974 \\
Mar07 & 0406800701 & 2007-03-06 & 37296 & 21903 & 2215 \\
Sep07 & 0504650201 & 2007-09-16 & 59014 & 34386 & 4117 \\
\enddata
\tablenotetext{a} {EPIC-pn}
\end{deluxetable*}

\medskip
\subsubsection{Pulse profiles}\label{1810pulse}

We started our analysis by making the simplest assumption about
the star surface thermal map, a uniform distribution at
temperature $T$. Lightcurves were then computed in the total, soft
and hard energy band for all the models in the archive. Once the
model closest to each observation (and in each band) was
found through the PCA, we used it as the starting point
for a fit performed using an IDL script based on the minimization
routine {\tt mpcurvefit.pro}. Our fitting function has six
free parameters, because, in addition to the twist angle, the
temperature, the electron velocity, the angles $\chi$ and $\xi$,
we have to include an initial phase to account for the
indetermination in the position of the pulse peak. Since it is not
possible to compute ``on the fly'' the pulse profile for a set of
parameters different from those contained in the archive,
lightcurves during the minimization process were obtained from
those in the archive using a linear interpolation in the parameter
space.

\begin{figure}
\includegraphics[width=3.3in,angle=0]{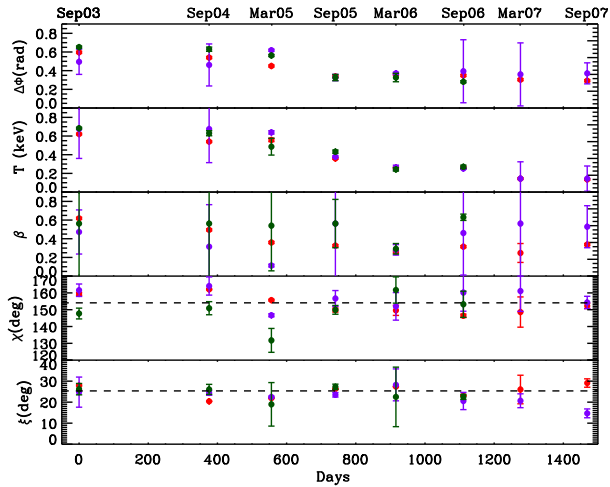}
\caption{Parameters evolution for XTE J1810-197, 
uniform surface temperature; results refer to the total (red dots), soft
(blue dots) and hard (green dots) energy bands.
Parameter errors are
calculated by the  minimization routine {\tt mpcurvefit.pro},
and are at $1\sigma$. Time is computed
starting from the September 2003 observation. \label{param_1810_1T}}
\end{figure}

In this way we obtained a fair agreement with the observed
pulse profiles ($\chi^2 \leq 1.12$ in five out of eight observations; see Table \ref{1810_comp_csq}),
and the values of the physical parameters
($\Delta\phi_{N-S}$, $\beta$, $T$) turn out to be the same (to within
the errors) for a given epoch among the different energy bands, as
it needs to be. Moreover, the evolution of the twist angle and of
the surface temperature follows a trend in which both quantities
decrease in time as the outburst declines.
This is expected if the outburst results from a sudden
change in the NS magnetic structure, producing both a heating of
the star surface layers and a twisting of the magnetosphere which
then dies away \citep{thompson02,belob09}. However, the model is
not acceptable since we found that the geometrical angles $\chi$
and $\xi$ change significantly from one observation to another,
and even for the same observation in the different energy bands
(see Fig. \ref{param_1810_1T} where the parameter evolution is
shown for the three energy bands). The analysis of the hard band was not carried
out after September 2006, because in both
the 2007 observations photons with energy $>2$ keV are only a few and, as a consequence,
lightcurves are affected by large uncertainties.

This shortcoming is most probably due to our oversimplifying
assumption about the NS thermal map. In fact, it was shown
that in the post-outburst phase the surface temperature
distribution of XTE J1810-197 is complex and changes in time
\cite[][although a different emission model was assumed in these
investigations]{perna08, bernardini09}. In order to check this, we
tried different configurations, starting with a
two-temperature map: a hot cap centered on one magnetic pole with
the rest of the surface at a constant, cooler temperature. In this
picture both temperatures as well as the emitting areas, are allowed
to vary in time.  While applying  the one-temperature model we
found that both the 2007 observations were reasonably well
reproduced with a value of the (uniform) surface temperature
$\sim 0.15 \ {\rm keV}$, comparable to the quiescent one \cite[see
also][]{bernardini09}. In order to check if this fit can be
further refined, we started from the September 2007 observation,
freezing the colder temperature at $T_c = 0.15 \ {\rm keV}$, and
letting the hot cap temperature $T_h$ free to vary. Since the cap
area $A_h$ is not known a priori, nor it can be treated as a free
parameter in our minimization scheme, we tried several values of
$A_h$, corresponding to one up to eight patches of our surface
grid (this means that $A_h$ is $n/32$ of the star surface,
with $n=1,\,\ldots,\, 8$). The best-estimate emitting area was
then taken as the one giving the lowest reduced $\chi^2$ for the
fit in the different trials. We verified that in all cases the
same value of the cap area produces the minimum $\chi^2$ in all
energy bands. Independent of the emitting area chosen, we always
found for $T_h$ a value compatible with $\sim 0.15 \ {\rm keV}$
for both the September and March 2007 observations.

One can then conclude that, for these two epochs, the entire star is
radiating at the same temperature, or if a hot cap exists, its
area is smaller than $\sim 3\, \%$ of the star surface (the size of
our surface grid resolution). For these epochs we report in
table \ref{1810_param_2T} the values of the cold temperature
obtained by using the single temperature scenario. We note
that the temperature value is at the border of our grid of
parameter values, so that, strictly speaking, it should be regarded as
an upper limit on $T_c$. However, we verified that the $\chi^2$
steeply grows when $T_c$ increases above $0.15-0.16\, {\rm keV}$.
Although there is no guarantee that the same is true when $T_c$
decreases, in the following we assume that $T_c \sim 0.15\,
{\rm keV}$ is a satisfactory estimate for the uniform temperature
at these epochs.

We then proceeded backwards in time, from September 2006 till
September 2003. Again, the cooler temperature is kept fixed while
several values of $A_h$ are tried. However, to account for the
possibility that also $T_c$ varies, we repeated the calculation
for $T_c = 0.15,\, 0.20, \, 0.25, \, 0.30 \ {\rm keV}$, looking
for the pair ($T_c, \, A_h$) which gives the lowest $\chi^2$.
Results are summarized in table \ref{1810_param_2T}. Although the
fits improve with respect to the one-temperature model (see
table~\ref{1810_comp_csq}), the two geometrical angles still
change from one observation to another and also across different
bands at the same epoch.

\begin{deluxetable*}{lcccccccc}
\tablecolumns{9} \tablewidth{0pc} \tablecaption{XTE
J1810-197 parameters and thermal map (two-temperature
model)\tablenotemark{a} \label{1810_param_2T}}
%\tablenum{1}
\tablehead{
\colhead{Epoch} &
\colhead{$\Delta\phi_{N-S}$} &
\colhead{$\beta$} &
\colhead{$\xi \ (^{\circ})$} &
\colhead{$\chi \ (^{\circ})$} &
\colhead{$T_{h}$ (keV)} &
\colhead{$A_{h} (\%)$} &
\colhead{$T_{c}$ (keV)} &
\colhead{$A_{c} (\%)$}
}
\startdata
Sep03 & $ 0.70 \pm 0.01 $ & $ 0.80 \pm 0.01 $ & $ 22.7 \pm 0.5 $ & $ 144.2 \pm 0.6 $ & $ 0.71 \pm 0.01 $ & $ 25.0 \pm 3.1 $ & $ 0.30 $ & $ 75.0 \pm 3.1 $\\
Sep04 & $ 0.67 \pm 0.01 $ & $ 0.62 \pm 0.02 $ & $ 20.7 \pm 1.0 $ & $ 158.2 \pm 0.2 $ & $ 0.55 \pm 0.01 $ & $ 18.7 \pm 3.1 $ & $ 0.30 $ & $81.3 \pm 3.1 $\\
Mar05 & $ 0.61 \pm 0.01 $ & $ 0.49 \pm 0.01 $ & $ 21.6 \pm 0.4 $ & $ 147.1 \pm 0.6 $ & $ 0.67 \pm 0.01 $ & $ 12.5 \pm 3.1 $ & $ 0.25 $ & $ 87.5 \pm 3.1 $\\
Sep05 & $ 0.47 \pm 0.01 $ & $ 0.53 \pm 0.05 $ & $ 23.0 \pm 0.1 $ & $ 159.0 \pm 1.1 $ & $ 0.42 \pm 0.01 $ & $ 9.4 \pm 3.1 $ & $  0.25  $ & $ 90.6 \pm 3.1 $\\
Mar06 & $ 0.49 \pm 0.01 $ & $ 0.50 \pm 0.11 $ & $ 23.5 \pm 0.4 $ & $ 149.4 \pm 3.5 $ & $ 0.28 \pm 0.01 $ & $ 6.2 \pm 3.1 $ & $  0.15  $ & $ 93.8 \pm 3.1 $\\
Sep06 & $ 0.43 \pm 0.01 $ & $ 0.71 \pm 0.16 $ & $ 21.4 \pm 0.3 $ & $ 155.7 \pm 1.7 $ & $ 0.28 \pm 0.01 $ & $ 3.1 \pm 3.1 $ & $  0.15  $ & $ 96.9 \pm 3.1 $\\
Mar07 & $ 0.45 \pm 0.01 $ & $ 0.6 \pm 0.01 $ & $ 29.8 \pm 0.1 $ & $ 162.8 \pm 0.1 $ & $ - $ & $ - $ & $ 0.16 \pm 0.01 $ & $ 100.0 $\\
Sep07 & $ 0.48 \pm 0.01 $ & $ 0.70 \pm 0.08 $ & $ 22.4 \pm 0.1 $ & $ 163.0 \pm 1.7 $ & $ - $ & $ - $ & $ 0.15 \pm 0.01 $ & $ 100.0 $\\
\enddata
\tablenotetext{a}{Total energy band; parameters with no reported
errors are fixed. Parameter errors are calculated by the
minimization routine {\tt mpcurvefit.pro}, and are at $1\sigma$.
Errors on the area correspond to the smallest patch of the grid.}
\end{deluxetable*}

In order to reproduce more accurately the star thermal map, we
generated a new model archive, increasing the number of surface
patches to $N_{\Theta} \times N_{\Phi} = 50 \times 4 = 200$. The
temperature, electron velocity and twist angle are in the range
$0.15 \ {\rm keV}\leq T \leq 0.9 \ {\rm keV}$ (step 0.15 keV),
$0.1 \leq \beta \leq 0.9$ (step 0.2) and $0.4\, {\rm rad}\leq
\Delta\phi_{N-S} \leq 1.2\, {\rm rad}$ (step 0.2 rad), respectively. We then
assumed that the star surface is divided into three zones: a hot
cap at temperature $T_{h}$, a concentric warm corona at $T_{w}$
and the remaining part of the neutron star surface at a cooler
temperature, $T_{c}$.  Again, we began our analysis from the 2007
observations, fixing $T_c = 0.15 \ {\rm keV}$, and searching for
the value of the warm temperature $T_w$. Every fit was
repeated for twelve values of the emitting area $A_w = 0.5\, \%,\,
1\, \%,\, 2\, \%, 4\, \%,\,\ldots,\, 20\, \%$ the total surface.
We found that the reduced $\chi^2$ improves with the addition of a
warm cap at $T_w \sim 0.3 \ {\rm keV}$, accounting for $0.5\, \%$
of the neutron star surface (see table \ref{1810_parameters}). We
stress that this value is below the resolution of our previous
grid, so the two results are consistent with each other.

We then considered the
two 2006 observations; in the two-temperature model based on
the previous archive, these were reasonably reproduced with $T_c=0.15 \
{\rm keV}$ and $T_h \sim 0.3 \ {\rm keV}$ (note that $T_h$ for the
two-temperature model corresponds to $T_w$ in the present case).
For these two observations we repeated the fit, fixing $T_c$ at
$0.15 \ {\rm keV}$ while leaving $T_w$ free to vary. The size of
the emitting area was estimated by following the same procedure
discussed above. We found an almost constant value, $T_w \sim 0.3
\ {\rm keV} $, between March 2006 and September 2007, while the
emitting area decreases in time. Also, we found no need for a
further component at $T_h$ at these epochs. On the other hand,
results for the two-temperature case (see table
\ref{1810_param_2T}) show the presence of a component with
temperature higher than $0.3 \ {\rm keV}$, in the period between
September 2003 and September 2005 (while the cooler one varies
between $0.25$ and $0.30 \ {\rm keV}$). It is tempting to
associate this to a transient hot cap that appears only in the
first period after the outburst, superimposed to the other,
longer-lived emitting zones.

\begin{figure*} 
\includegraphics[width=3.3in,angle=0]{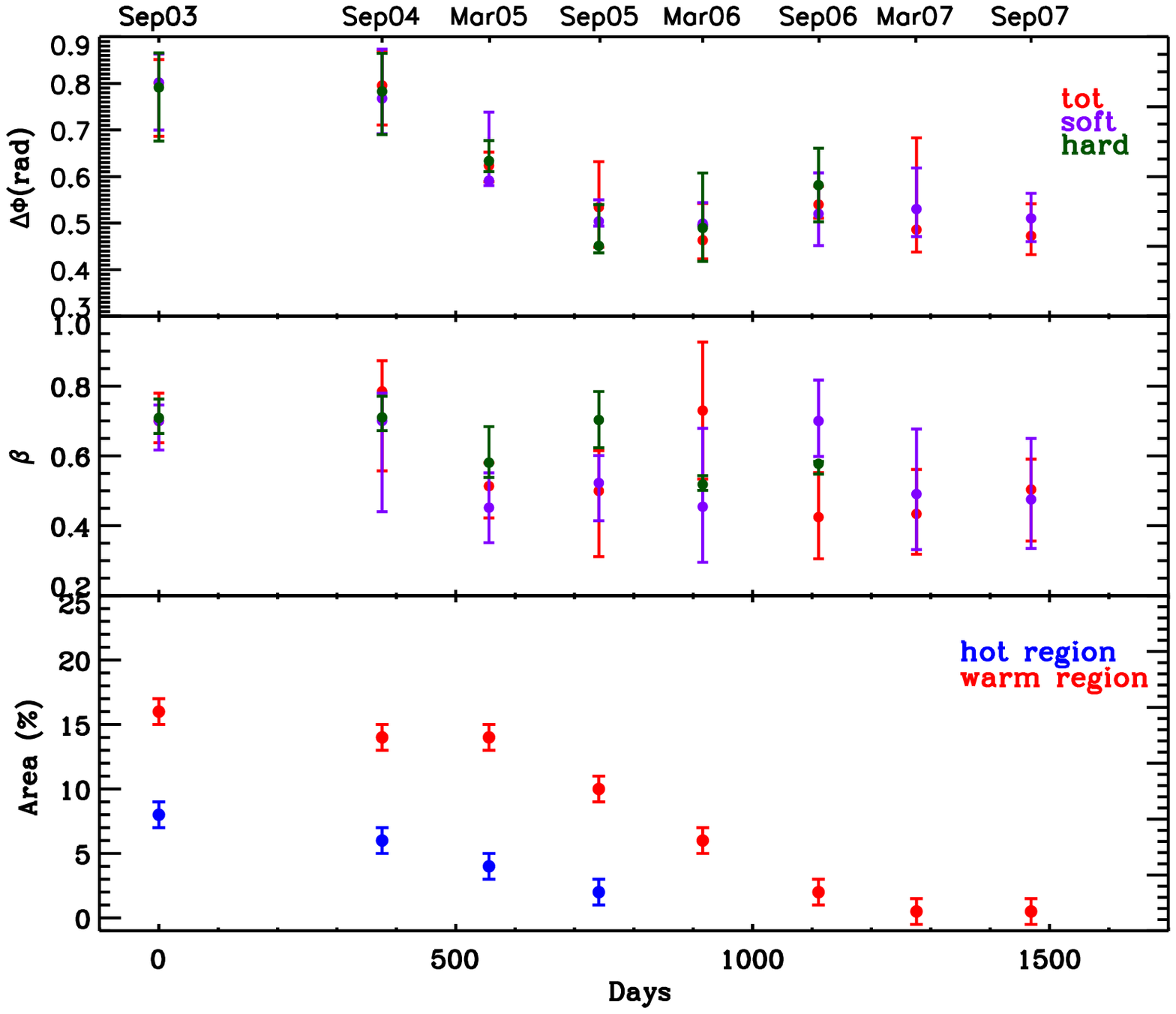} 
\includegraphics[width=3.3in,angle=0]{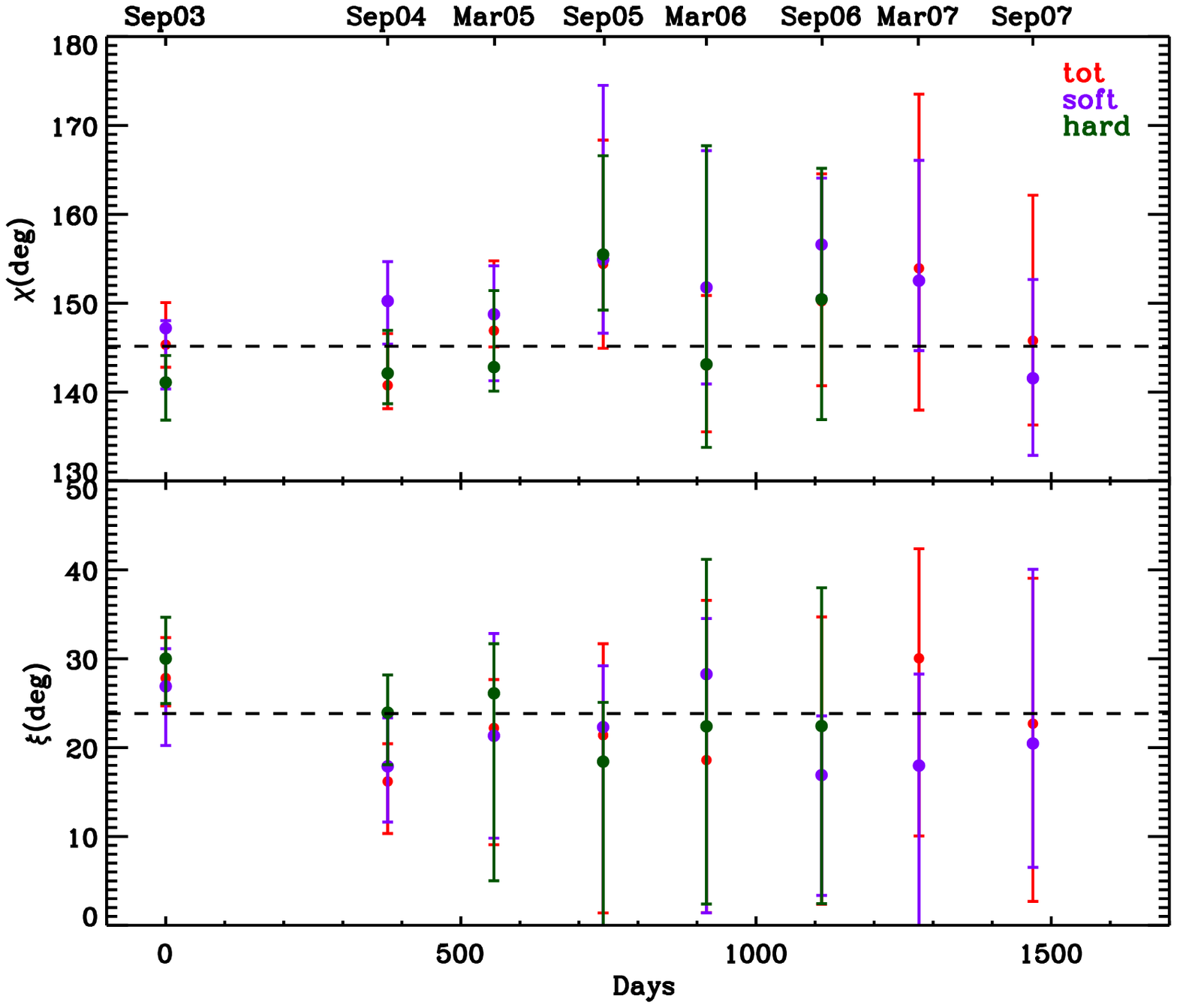} \
\caption{Parameters 
evolution for XTE J1810-197, three temperature model. Left (from top to 
bottom): twist angle ($\Delta \phi$), bulk velocity ($\beta$) and area of 
the different emitting regions. Right: the two geometrical angles, $\chi$ 
and $\xi$. Details as in fig. \ref{param_1810_1T}. 
\label{param_1810}} 
\end{figure*}

To test this possibility, we re-fitted the
first four observations by fixing the coldest temperature at $T_c
= 0.15 \ {\rm keV}$, the warmer one at $T_w = 0.3 \ {\rm keV}$,
and leaving only the hotter temperature free to vary. For each
observation the pulse profile fits were computed for every
combination of $A_h$ and $A_w$ chosen among the twelve values in
the range $0.5\, \%$-- $20\,\%$ introduced before, and looking for
the minimum of the reduced $\chi^2$. Results of the lightcurve
fitting at different epochs are listed in table
\ref{1810_parameters} and shown in Fig. \ref{param_1810}, while
the reduced $\chi^2$ for the three thermal
distributions is reported in table \ref{1810_comp_csq}.

\begin{deluxetable*}{lccccccccc}
\tablecolumns{10} \tablewidth{0pc}\tablecaption{XTE J1810-197
parameters and thermal map (three-temperature
model)\tablenotemark{a} \label{1810_parameters}}
%\tablenum{1}
\tablehead{
\colhead{Epoch} &
\colhead{$\Delta\phi_{N-S}$} &
\colhead{$\beta$} &
\colhead{$\xi \ (^{\circ})$} &
\colhead{$\chi \ (^{\circ})$} &
\colhead{$T_h$ (keV)} &
\colhead{$T_w$ (keV)} &
\colhead{$T_c$ (keV)} &
\colhead{$A_h (\%)$} &
\colhead{$A_w (\%)$}
}
\startdata
Sep 03 & $0.80^{+0.05}_{-0.11}$ & $0.70^{+0.08}_{-0.06}$ & $27.8^{+4.6}_{-3.1}$ & $145.3^{+4.7}_{-2.5}$  & $0.62^{+0.14}_{- 0.14}$ & $0.30$ & $0.15$ & $8.\pm0.5$ & $16.\pm0.5$\\
Sep 04 & $0.79^{+0.07}_{-0.08}$ & $0.78^{+0.09}_{-0.23}$ & $16.2^{+4.2}_{-5.9}$ & $140.8^{+5.8}_{-2.6}$  & $0.49^{+0.03}_{-0.22}$ & $0.30$ & $0.15$ &$6.\pm0.5$ &  $14.\pm0.5$\\
Mar 05 & $0.62^{+0.03}_{-0.03}$ & $0.51^{+0.07}_{-0.09}$ & $22.2^{+5.4}_{-13.1}$ & $146.9^{+7.9}_{-1.8}$  & $0.49^{+0.01}_{-0.04}$ & $0.30$ &$0.15$ & $4.\pm0.5$ & $14.\pm0.5$\\
Sep 05 & $0.53^{+0.10}_{-0.09}$ & $0.50^{+0.11}_{-0.19}$ & $21.4^{+10.3}_{-20.0}$ & $154.4^{+13.9}_{-9.5}$ & $0.52^{-}_{-}$ & $0.30$ & $0.15$ &$2.\pm0.5$ & $10.\pm0.5$\\
Mar 06 & $0.46^{+0.08}_{-0.04}$ & $0.73^{+0.20}_{0.20}$ & $18.5^{18.0}_{-17.2}$ & $143.0^{+7.9}_{-7.5}$  & - & $0.29^{+0.17}_{-0.03}$ & $0.15$ & - & $6.\pm0.5$ \\
Sep 06 & $0.54^{+0.04}_{-0.03}$ & $0.42^{+0.13}_{-0.12}$ & $22.4^{+12.3}_{-20.0}$ & $150.2^{+14.3}_{-9.5}$ & - & $0.30^{-}_{-}$ & $0.15$ & - & $2.\pm0.5$ \\
Mar 07 & $0.49^{+0.20}_{-0.05}$ & $0.43^{+0.13}_{-0.12}$ & $30.0^{+12.3}_{-20.0}$ & $153.9^{+19.6}_{-16.0}$  & - & $0.29^{-}_{-}$ & $0.15$ & - & $0.5\pm0.5$\\
Sep 07 & $0.47^{+0.07}_{-0.04}$ & $0.50^{+0.09}_{-0.15}$ & $22.7^{+16.4}_{-20.0}$ & $145.8^{+16.4}_{-9.5}$ & - & $0.31^{-}_{-}$ & $0.15$ & - & $0.5\pm0.5$ \\
\enddata
\tablenotetext{a}{Total energy band; parameters with no reported
errors are fixed. Errors are computed from the $\chi^2$ curve (see text
for details) and are at $1\sigma$. No errors are reported when
they could not be calculated (flat $\chi^2$ curves) and
errors on the area have the same meaning as in Tab. \ref{1810_param_2T}.}
\end{deluxetable*}

\begin{deluxetable}{lccccc}
\tablecolumns{6}
\tablewidth{0pc}
\tablecaption{Reduced $\chi^2$ for XTE J1810-197 \tablenotemark{a}
\label{1810_comp_csq}}
%\tablenum{1}

\tablehead{
\colhead{Epoch} &
\colhead{$\chi^2_{red}$} &
\colhead{$\chi^2_{red}$} &
\colhead{$\chi^2_{red}$} &
\colhead{$\chi^2_{red}$} &
\colhead{$T$} \\
\noalign{\smallskip}
\colhead{}&
\colhead{(1T)} &
\colhead{(2T)} &
\colhead{(3T)} &
\colhead{(XSPEC)} &
\colhead{(keV)} 
}

\startdata
Sep 03 & 1.72 & 1.58 & 0.12 & 1.22 & - \\
Sep 04 & 0.66 & 0.42 & 0.36 & 1.93 & - \\
Mar 05 & 1.02 & 0.98 & 0.79 & 1.50 & - \\
Sep 05 & 1.06 & 0.40 & 0.39 & 1.52 & $0.53_{-0.06}^{+0.07}$  \\
Mar 06 & 2.94 & 1.70 & 1.25 & 1.34 & - \\
Sep 06 & 0.94 & 0.38 & 0.35 & 1.36 &  $0.31_{-0.01}^{+0.03}$ \\
Mar 07 & 2.88 & 2.88 & 2.37 & 1.08 & $0.29_{-0.02}^{+0.04}$ \\
Sep 07 & 1.12 & 1.12 & 0.96 & 1.29 & $0.31_{-0.01}^{+0.01}$  \\
\enddata
\tablenotetext{a}{First three columns:
reduced $\chi^2$ obtained from the lightcurves fitting
for total the energy band (results for the other two bands are similar).
Last two columns: reduced $\chi^2$ obtained from the spectral fitting
in {\tt XSPEC}, and corresponding temperatures.
The temperature was left free to vary only at those epochs and for those
components for which the lightcurve analysis did not produce an unique
value. Errors for the temperature are at $1 \sigma$.}
\end{deluxetable}

A worry may arise whether the best-fitting values obtained
from the minimization routine correspond indeed to absolute
minima of the reduced $\chi^2$. In order to check this, and
visually inspect the shape of the $\chi^2$ curve close to the
solution, we computed and plotted the reduced $\chi^2$ leaving, in
turn, only one parameter free and freezing the remaining five at
their best-fit values. This also allowed us to obtain a more
reliable estimate of the parameter errors which were computed
by looking, as usual, for the parameter change which corresponds to a
$1\sigma$ confidence level (and reported in table
\ref{1810_parameters}).

\begin{figure}
\includegraphics[width=3.3in]{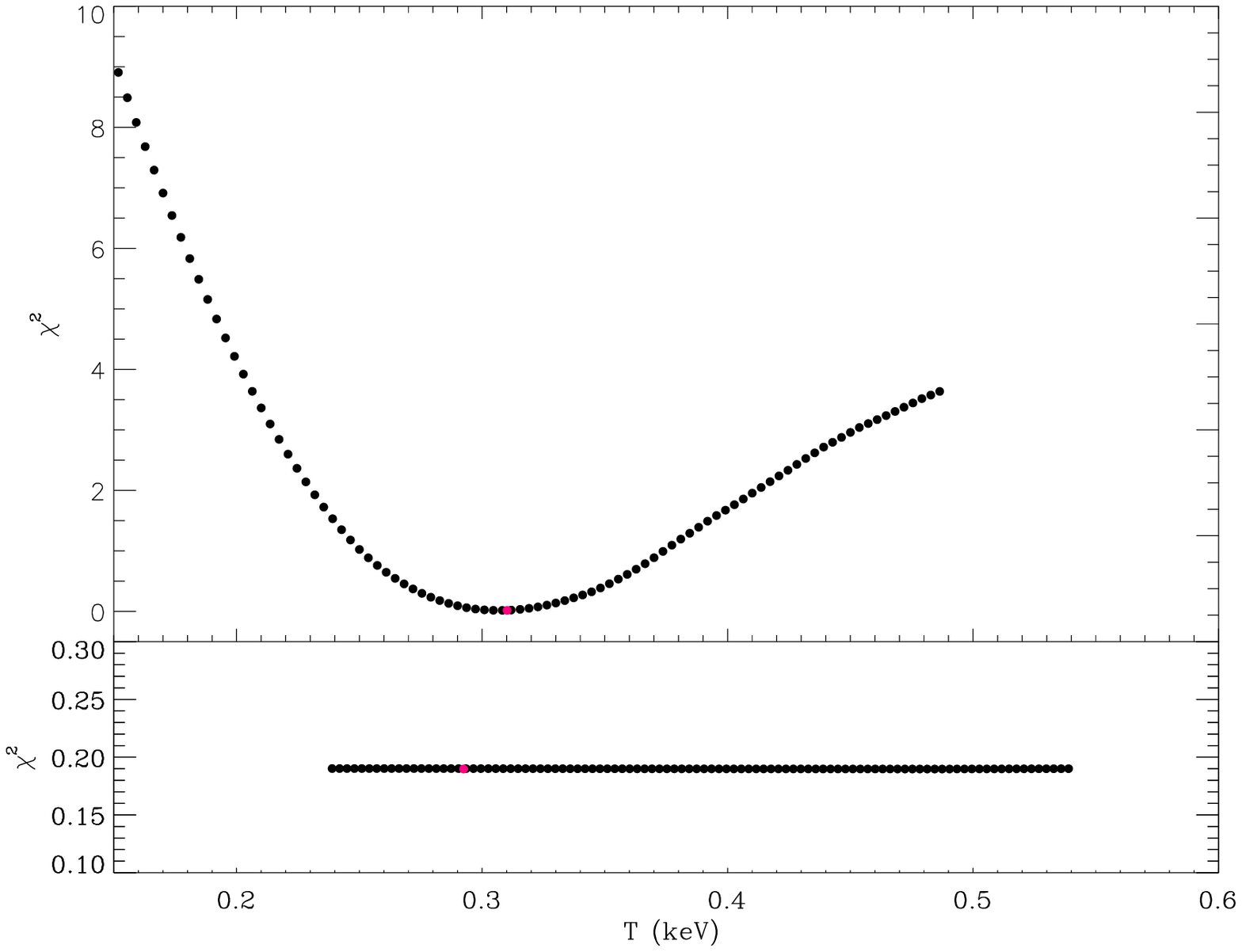}
\caption{Two examples of the different behavior of the reduced
$\chi^2$ for the warm temperature. The top curve refers to the March 2006 observation (soft
band) and exhibits a well-defined minimum. The bottom curve (September
2006
observation, total band) is so flat to make it impossible to gauge the
best-fitting value and its errors.
\label{csq_ex}}
\end{figure}

We found that all values obtained with the {\tt mpcurvefit.pro} routine indeed
correspond to minima of the reduced $\chi^2$ curve, with the exception
of the temperature(s), for which there are
observations (or energy bands) with very flat $\chi^2$
curves (see Fig. \ref{csq_ex}). In particular,
for the September 2005 observation the curve
obtained varying $T_h$ is flat in all the three energy bands. Also
the curves relative to $T_w$ for the September 2006, March 2007
and September 2007 observations have the
same problem. This can be understood by noting that in all these
observations the size of the hot/warm region accounts for only $<2\, \%$
of the total neutron star surface: temperature changes
in such a small emitting area can hardly influence the fit.
In addition, for the March 2005 and March 2006 observations the reduced
$\chi^2$ curve relative to one of the temperatures is flat, but this
occurs only for one of the three energy bands. The first case concerns
the hot temperature and the soft band, the second the warm temperature
and the hard band. As we discussed above, when the hot (warm)
area shrinks it affects little the pulse profile; this shows
up first in the energy band in which its emission contributes less, i.e.
the soft (hard) band.

Given these findings, we concluded that lightcurve analysis by itself is
unable to yield an unique temperature value for the
September 2005, the September 2006, and both the 2007 observations. On
the other hand,  spectral analysis is more sensitive to temperature
variations, so that it is possible to infer a temperature value
also in these cases. As it will be discussed in the next section,
by combining the two techniques we can remove most of the
uncertainties and validate the three temperature model presented
so far (see sec. \ref{1810spec} for details).

There are several physical implications than can be drawn
from our model. The TAXP is seen at an angle $\chi =
148^{\circ +7}_{-9}$ with respect to the spin axis. The
misalignment between the spin axis and the magnetic axis is $\xi =
23^{\circ +15}_{-11}$. These values of the two angles, and the corresponding
errors, are calculated from the weighted average in the three energy bands. To get a quantitative
confirmation that $\chi$ and $\xi$ do not change in time, we fitted
a constant through the values of each angle as derived from
the lightcurves fitting at the different epochs and found that the null hypothesis probability
is $<1\, \%$. We note that, formally, the misalignment between the spin and the magnetic axis is
compatible with being zero at the $3 \sigma$ level. Low values of $\xi$ produce, however, models
with pulsed fractions quite smaller than the observed ones and, despite $\xi\sim 0$ might be still statistically
acceptable, we regard this possibility as unlikely because the amplitude is the main feature which characterizes
the pulse, as the PCA shows (see~\S\ref{pca}, the first principal component, $Z_1$, is, in fact, directly
related to the amplitude).

It emerges a scenario in which, before the outburst, the NS surface
radiates uniformly at a temperature $T_{c} \sim 0.15 \ {\rm keV}$.
Soon after the burst the thermal map of XTE J1810-197
substantially changes. The region around the magnetic north pole
is heated, reaches a temperature of $\sim 0.7 \ {\rm keV}$ and
covers an area $\sim 8\%$ of the total star surface. This hot
spot is surrounded by a warmer corona at $\sim 0.3$ keV, that
covers a further $\sim 16\%$ of the surface. During the subsequent
evolution, the hot cap decreases in size and temperature until
the March 2006 observation, when it becomes too small and cold to
be distinguished from the surrounding warm corona. The warm region
remains almost constant until September 2005, then decreases in
size, and becomes a cap in March 2006, following the hot spot
disappearance. In September 2007 (our last observation for XTE
J1810-197) the warm cap is still visible, even if its area is down
to only $\sim 0.5\%$ of the total. The twist angle is highest at
the beginning of the outburst (September 2003) and then steadily
decreases until it reaches a more or less constant value around
September 2005. The electron velocity does not show large
variations in time and stays about constant at $\beta\sim
0.5$.

\begin{figure}
\includegraphics[width=3.3in]{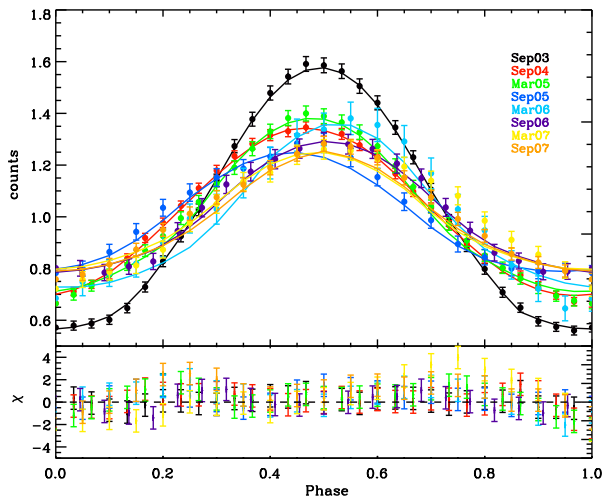}
\caption{Synthetic (3T model) and observed pulse profiles for XTE J1810-197 in
the total energy band. Solid lines represent the best-fitting model, dots the observed
lightcurves. Initial phases are arbitrary. The lower panel shows the residuals.
\label{1810_lc}}
\end{figure}

\begin{figure}
\includegraphics[width=3.3in]{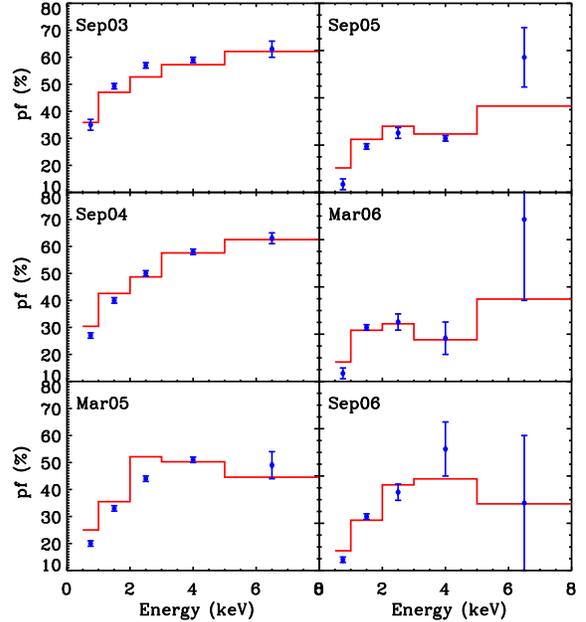}
\caption{The variation of XTE J1810-197 pulsed fraction with energy at
different epochs. The red line refers
to the model, blue dots to observations (errors are at $1\sigma$).
\label{pf_1810}}
\end{figure}

Synthetic and observed lightcurves (in the total band) are shown in
Fig \ref{1810_lc}, together with the fit residuals. We note that
the residuals exhibit a well-defined, oscillatory pattern at all epochs.
In our scenario, this can be possibly associated to a more complicated
thermal map, of which our 3T model is a first-order approximation (e.g.
non-circular shape of the hotter regions, off-centering of the hot and warm
areas). However, as discussed in some more detail in \S~\ref{discus}, no further
refinement of the surface thermal map will be attempted here.
Since XTE J1810-197 pulse profiles are fairly sinusoidal, we can compute
the pulsed fraction and its evolution in time at different energies.
The comparison of model results with data is shown in Fig. \ref{pf_1810}.

\newpage

\subsubsection{Spectra}\label{1810spec}

\begin{figure}
\includegraphics[width=3.3in]{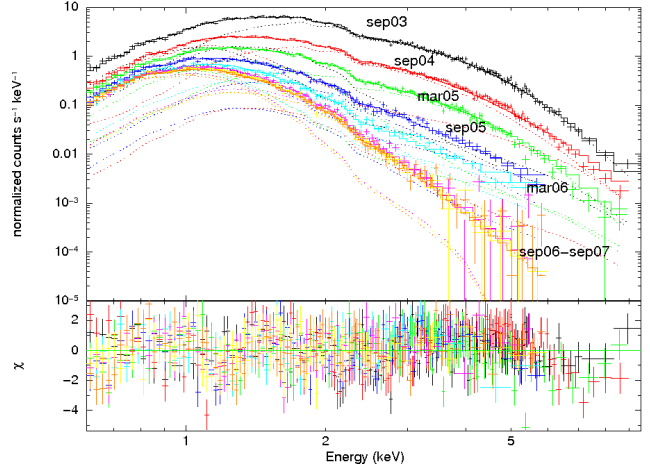}
\caption{Spectral evolution in the eight {\it XMM-Newton} observations of
XTE J1810-197. Solid lines represent the model, while dotted lines refer to
the single {\tt ntzang} components (see text for details). Residuals
are shown in the lower panel. \label{1810_spectra}}
\end{figure}

In order to verify if the thermal map inferred from the pulse
profile fits is reasonable, and in order to remove the
uncertainties in the value of the temperature at certain epochs
(see \S~\ref{1810pulse}), we examined the source spectra. The goal
is to check if the parameters derived from our lightcurve analysis
(twist angle, bulk velocity, size and temperature of the three
emitting areas, and the angles $\chi$, $\xi$) can also
reproduce the spectral evolution of XTE J1810-197 during the
outburst decay. To this end, we used the {\tt ntzang} model
that was implemented in {\tt XSPEC} by \citet[][the model is
not currently available in the public library, but it can be obtained from
the authors upon request]{nobili08}. The {\tt ntzang XSPEC} model has the same
free parameters as those used in our fits of the pulse profiles. In addition
it contains the normalization and the column density. We caveat that, since
this {\tt XSPEC} model was created by assuming that the entire
star surface emits at uniform temperature, strictly speaking is
not directly suited to the present case. As an approximation, we
fitted the spectra by adding together three (absorbed) {\tt ntzang}
models, each associated to one of the three thermal components, at
temperatures $T_{h}$, $T_{w}$ and $T_{c}$, respectively. At each
epoch the fit was performed by freezing $\Delta\phi_{N-S}$,
$\beta$, $T$, $\chi$ and $\xi$ at the values derived from the fit
of the lightcurve in the total energy band (see
\S\ref{1810pulse}), while the three model normalizations (which
are related to the emitting areas) were left free to vary. We also
required that the column density, $N_H$, is the same for all the
three spectral components and for all epochs. Since for the
September 2005, the September 2006 and both the 2007 observations the
lightcurve analysis did not return an unique value for the
hotter temperature, we also left this parameter free to vary in
these four observations.
 In all these cases, we found that the fit
converges to a value of the temperature close to the best-fitting
value obtained from the lightcurve analysis (see
table~\ref{1810_comp_csq}). Moreover, the reduced $\chi^2$
significantly worsens by varying the temperature, meaning that the
spectra are much more sensitive to the presence of these
components. Results are shown in Fig. \ref{1810_spectra}, while
the reduced $\chi^2$ for the fits at the various epochs are
reported in Tab. \ref{1810_comp_csq}. The value of the column
density is found to be $N_H = (7.73  \pm 0.50) \times 10^{21} \
{\rm cm^{-2}}$, compatible at the $1.5 \sigma$ level with the one
obtained by \cite{bernardini09} with the 3 BB model, $N_H = (6.3
\pm 0.5) \times 10^ {21} \ {\rm cm^{-2}}$. We remark that, in
assessing the goodness of the fits, only the normalizations of the
three components (plus $N_H$) are free to vary; all the other
model parameters are frozen at the best values obtained from the
pulse profile analysis. Under these conditions, we regard the
agreement of our model with observed spectra as quite
satisfactory. We note that the presence of systematic residuals at high energies (above
7--8 keV) may be hinted in the fits of the three earlier observations (see Fig. \ref{1810_spectra}).
As discussed by \cite{bernardini09} they may be related to a harder spectral component which
is however only marginally significant ($3.2\sigma$ confidence level) and quite unconstrained.
Given that the high-energy residuals are comparable to (or smaller than) those of the
3 BB model used by \cite{bernardini09}, we conclude that a hard tail is not significant  also
in our modelling and we did not attempt to include it in our fits.

We checked how the reduced $\chi^2$ for the spectral fit changes when the (frozen) parameters are varied
within $\sim 2\sigma$ from their best-fit value (as from the pulse fitting). This has been done
changing one parameter at a time. We found that indeed the $\chi^2$ increases quite smoothly in response
to the change of each parameter, with the exception of $\chi$ and $\xi$. This is not surprising,
since we knew already that the spectrum is not much sensitive to the geometry. We also
tried a fit leaving all the parameters free, apart from the two geometrical angles which were
held fixed at their best-fit values. The fit returns parameter values which are the same, within
the errors, as those derived from the pulse fitting and comparable values of the reduced $\chi^2$, implying that
the solution we presented is indeed a global $\chi^2$ minimum. The same procedure and the same conclusions
hold also in the case of CXOU 164710.2-455216 (see \S~\ref{1647spectra}).

\subsection{CXOU J164710.2-455216}\label{1647anal}

Having verified that our model can provide a reasonable interpretation
for the post-outburst timing and spectral evolution of TAXP prototype
XTE J1810-197, we applied it to CXOU J164710.2-455216, the other transient
AXP for which a large enough number of {\it XMM-Newton} observations
covering the outburst decay are available (see table \ref{CXOUobs}
for details). During September 2006 the pn and MOS cameras were set in full window
imaging mode with a thick filter (time resolution =
$73.3\times 10^{-3}$ s and 2.6 s for the pn and MOS, respectively),
while all other observations were in a large and small window imaging
mode with a medium filter (time resolution = $4.76\times 10^{-2}$ s and
0.3 s for the pn and MOS, respectively). To extract more than 90\% of
the source counts, we accumulated a one-dimensional image and fitted the
1D photon distribution with a Gaussian. Then, we extracted the source
photons from a circular region of radius 40\arcsec (smaller than the
canonical 55\arcsec, corresponding to 90\% of the source photons, in
order to minimize the contamination from nearby sources in the Westerlund 1 cluster)
centered at the Gaussian centroid. The background
for the spectral analysis was obtained (within the same pn CCD where the
source lies and a different CCD for the MOS) from an annular region (inner and outer radii of 45\arcsec
and 65\arcsec , respectively) centered at the best source position. In
the timing analysis, the background was estimated from a circular region
of the same size as that of the source. EPIC-pn spectra were processed as in the case of XTE J1810-197 (see
\S\ref{1810anal}).

\begin{deluxetable*}{lccccc}
\tablecolumns{6}
\tablewidth{0pc}
\tablecaption{CXOU J164710.2-455216 {\it XMM-Newton} observations \tablenotemark{a}
\label{CXOUobs}}
%\tablenum{1}
\tablehead{\colhead{Label} &
\colhead{OBS ID} &
\colhead{Epoch} &
\colhead{Exposure time (s)} &
\colhead{total counts} &
\colhead{background counts}
}
\startdata
Sep 06 & 0311792001 & 2006-09-22 & 26780 & 56934 & 1709 \\
Feb 07 & 0410580601 & 2007-02-17 & 14740 &18734  & 1264 \\
Aug 07 & 0505290201 & 2007-08-19 & 16020 & 11710 & 2384 \\
Feb 08 & 0505290301 & 2008-02-15 & 9080 & 4618 & 1131 \\
Aug 08 & 0555350101 & 2008-08-20 & 26360 & 7357 & 1689 \\
Aug 09 & 0604380101 & 2009-08-24 & 33030 & 4974 & 1959 \\
\enddata
\tablenotetext{a}{EPIC-pn}
\end{deluxetable*}

\subsubsection{Pulse profiles}\label{1647pulse}

\begin{figure*}
\includegraphics[width=3.3in,angle=0]{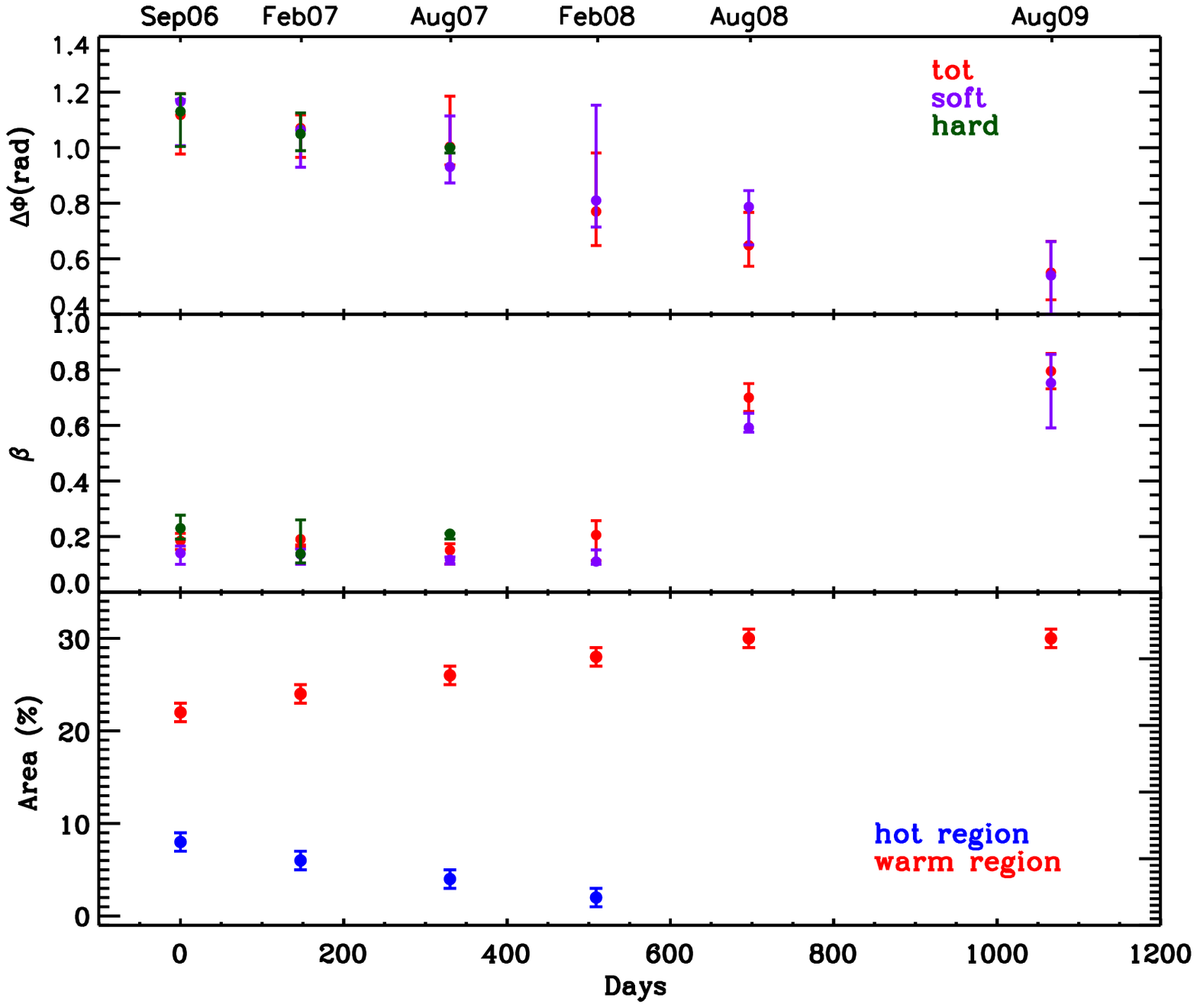}
\includegraphics[width=3.3in,angle=0]{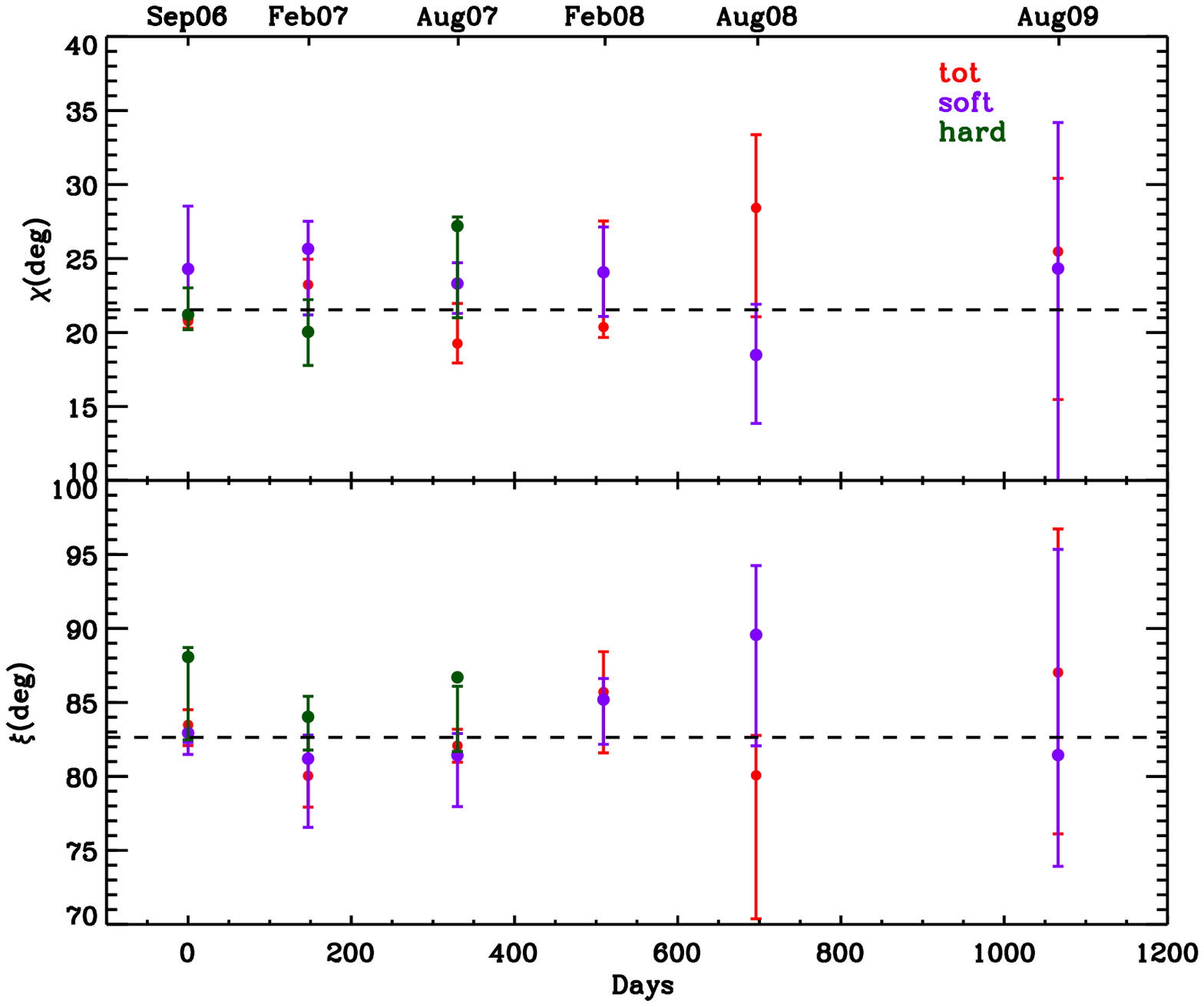}
\caption{Same as in Fig. \ref{param_1810} for the TAXP CXOU
J164710.2-455216; here time is computed starting from the September 2006
observation. \label{param_1647}}
\end{figure*}

The analysis of the pulse profiles of CXOU J164710.2-455216
follows closely that presented in \S \ref{1810pulse}. In
particular, we first tried a single temperature and then a
two-temperature model, encountering the same problems we found for
XTE J1810-197. Finally, we applied a three-zone thermal map and
this provided reasonable fits for the lightcurves, and the angles $\chi$ and $\xi$
were not found to vary in the same
observation for the different energy bands and for different
epochs. We did not attempt to fit the pulse profiles in the hard band after the
February 2007 observation because of the very few counts
at energies $> 2$ keV. As for XTE
J1810-197 we started the analysis from the last observation
(August 2009) assuming a thermal map comprising a hot cap centered
on the magnetic pole at temperature $T_h$, a concentric warm
corona at $T_w$ and the rest of the neutron star at the colder
temperature $T_c$. Every fit was repeated for ten values of
the hot cap area $A_h = 0.5\, \%,\, 1\, \%,\, 2\, \%, 4\,
\%,\,\ldots,\, 16\, \%$ (of the total surface) and for 20 values
of the warm corona area $A_w = 0.5\, \%,\, 1\, \%,\, 2\, \%, 4\,
\%,\,\ldots,\, 30\, \%$. Moreover, lightcurves fits were
iterated for two values of the cold temperature $T_c = 0.15, 0.30$
keV and also for two values of the warm temperature $T_w = 0.30,
0.45$ keV. The hotter temperature was left free to vary. We found that in the
last two observations, independent of the hot cap size, $T_h$ is
always $\sim 0.45$ keV, nearly indistinguishable from the
temperature of the warm corona obtained from the fit. We concluded
that, at least for our present surface grid resolution, in the
last two observations there are only two thermal components that
contribute to the emission, the cold and warm ones, and repeated the
fit leaving $T_w$ free to vary.
Results are
reported in table \ref{1647_parameters} and plotted in Fig.
\ref{param_1647}, while a comparison of the reduced $\chi^2$ for
the three thermal distributions is given in table
\ref{1647_comp_csq}. Errors listed in the tables have the same
meaning as in the case of XTE J1810-197. Again, when the spot at
$T_h$ becomes very small its temperature can not be determined
unambiguously.

\begin{deluxetable*}{lccccccccc}
\tablecolumns{11}
\tablewidth{0pc}
\tablecaption{CXOU J164710.2-455216 parameters and thermal map
(three-temperature model)\tablenotemark{a} \label{1647_parameters}}
\tablehead{
\colhead{Epoch} &
\colhead{$\Delta\phi_{N-S}$} &
\colhead{$\beta$} &
\colhead{$\xi \ (^{\circ})$} &
\colhead{$\chi \ (^{\circ})$} &
\colhead{$T_{h}$ (keV)} &
\colhead{$T_{w}$ (keV)} &
\colhead{$T_{c}$ (keV)} &
\colhead{$A_{h} (\%)$} &
\colhead{$A_{w} (\%)$}
}
\startdata
Sep 06 & $1.12^{+0.08}_{-0.14}$ & $0.18^{+0.03}_{-0.03}$ & $83.5^{+1.0}_{-1.4}$ & $20.8^{+0.1}_{-0.5}$  & $0.70^{+0.20}_{-0.11}$ & $0.45$ & $0.15$ & $8.\pm0.5$ & $22.\pm0.5$ \\
Feb 07 & $1.07^{+0.05}_{-0.10}$ & $0.19^{+0.02}_{-0.03}$ & $80.0^{+2.7}_{-2.1}$ & $23.2^{+1.7}_{-1.7}$ & $0.64^{+0.16}_{-0.06}$ & $0.45$ & $0.15$ &$6.\pm0.5$ &  $24.\pm0.5$\\
Aug 07 & $1.00^{+0.18}_{-0.06}$ & $0.15^{+0.02}_{-0.05}$ & $82.1^{+1.1}_{-1.1}$ & $19.2^{+2.7}_{-1.3}$ & $0.63^{+0.18}_{-0.10}$ & $0.45$ &$0.15$ & $4.\pm0.5$ & $26.\pm0.5$ \\
Feb 08 & $0.77^{+0.21}_{-0.12}$ & $0.20^{+0.05}_{-0.09}$ & $85.7^{+2.7}_{-4.1}$ & $20.4^{+7.2}_{-0.7}$ & $0.62^{-}_{-}$ & $0.45$ & $0.15$ &$2.\pm0.5$ & $28.\pm0.5$ \\
Aug 08 & $0.65^{+0.12}_{-0.07}$ & $0.70^{+0.05}_{-0.05}$ & $80.1^{+2.7}_{-9.7}$ & $28.4^{+4.9}_{-7.3}$ & - & $0.49^{+0.02}_{-0.05}$ & $0.15$ & - & $30.\pm0.5$ \\
Aug 09 & $0.55^{+0.11}_{-0.10}$ & $0.79^{+0.06}_{-0.06}$ & $87.0^{+9.7}_{-10.9}$ & $25.5^{+4.9}_{-10.}$ & - & $0.46^{+0.05}_{-0.05}$ & $0.15$ & - & $30.\pm0.5$ \\
\enddata
\tablenotetext{a}{Total energy band. Errors have the same meaning as in Tab.
\ref{1810_parameters}}
\end{deluxetable*}

According to our model, CXOU J164710.2-4552116 is viewed at an
angle $\chi = 23^{\circ +4}_{-3}$  with
respect to its spin axis.
The spin and the magnetic axes are almost orthogonal, $\xi =
84^{\circ +5}_{-3}$. This is a quite peculiar condition, and it
seems to be the only one capable of explaining
the characteristic three-peaked shape of the observed
lightcurves within the present model.
As for XTE J1810-197 values and errors for both angles are
calculated as the weighted average of parameters in the three energy bands.
Also in this case the probability that $\chi$ and
$\xi$ are not constant in time is $< 1\%$.

Soon after the burst the thermal map of CXOU J164710.2-455216 consists of
three regions at different  temperatures. The hottest region,
around the north magnetic pole, has a temperature $T_{h} \sim
0.7 \ {\rm keV}$, and its area is $\sim 8\%$ of the total. This
hot spot decreases in temperature and size as time elapses,
until February 2008. In August 2008 the hot cap becomes
so small in  size and its temperature so close to that of the warm
corona, that it is impossible to distinguish between the two regions.
The warm corona has a temperature of $\sim 0.45 \ {\rm keV}$,
which remains about constant during the three years of
observations. In this case the corona area slightly increases with time,
starting from $\sim 20\%$ and reaching $\sim 30\%$ of the NS
surface. The third region has a lower temperature $T_{c} \sim 0.15
\ {\rm keV}$ and its area remains constant at $\sim 70\%$  of the
total. The twist angle is $\sim 1.12$ rad soon after the burst, and it
decreases with time. There are hints that its decay is slower
until August 2007, then proceeds faster. The electron velocity is
about the same at all epoch ($\beta\sim 0.2$), apart from the last
two observations in which it strongly increases. This variation may
be related to the change in the pulse shape (from three-peaked to
single-peaked) and also to the increase of the pulsed fraction.

\begin{figure}
\includegraphics[width=3.3in]{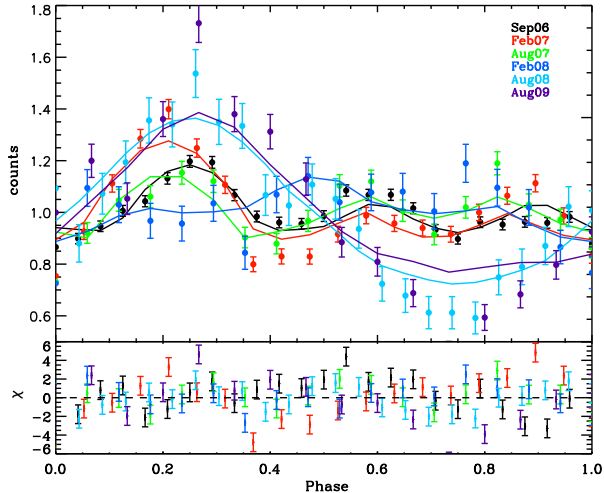}
\caption{Same as in Fig. \ref{1810_lc} for CXOU J164710.2-455216;
computed pulse profiles refer to the 3T model and initial phases are arbitrary. \label{1647_lc}}
\end{figure}

A comparison between the observed and model pulse profiles is
shown in Fig. \ref{1647_lc}. Because of the inherent complexity
and drastic time evolution of CXOU J164710.2-455216 lightcurves,
the agreement is not as good as for XTE J1810-197. The fact that
lightcurve fits return $\chi^2$ values not much higher than those
of XTE J1810-197 (compare Tab.~\ref{1810_comp_csq} and
\ref{1647_comp_csq}) reflects the larger uncertainties in the
phase-binned source counts. We also note that the errors on the geometrical
angles for CXOU J164710.2-455216 are smaller than those derived for
XTE J1810-197, despite the worst agreement (see Tab. \ref{1810_parameters} and 
\ref{1647_parameters}). This is most probably due to the different shapes
of the pulses in the two sources. Because of the very peculiar lightcurve
of CXOU J164710.2-455216, which can be reproduced by our model
only invoking a nearly orthogonal rotator, even small depatures of $\xi$ and $\chi$
from their best-fit values results in a rapid growth of the $\chi^2$. This does not
occur for the rather sinusoidal pulse of XTE J1810-197 since the model can produce
lightcurves of more or less the right shape in a wider range
of angles.

Besides being of limited use because
of the complex shape of the pulse, the pulsed fraction analysis 
was hindered by the lower count rate, especially at low energies
and was not pursued further for this source. As in XTE J1810-197,
we checked that the values obtained from the minimization routine
indeed correspond to minima of the reduced $\chi^2$s. Again we
froze five of the six parameters to the value obtained with the
{\tt mpcurvefit.pro} minimization routine, and calculated the
reduced $\chi^2$ around its minimum by varying the free parameter.
The procedure was repeated for all parameters and all observations
in the three energy bands. Again, for all parameters but the
temperature, results obtained with the {\tt mpcurvefit.pro}
routine indeed correspond to the minima of the reduced $\chi^2$
curve. There is one observation for which the $\chi^2$ curve
relative to the hot temperature is very flat for all the energy
bands. This is the August 2008 observation, for which the size of
the emitting area accounts for just $2 \%$ of the total neutron
star surface. As in XTE J1810-197, we conclude that the fit is not
very sensitive to the temperature variation for very small
emitting areas. On the other hand, like in the previous case, it
was possible to infer a value for the August 2008 hot temperature
using the spectral analysis (see sect. \ref{1647spectra}).

\begin{deluxetable}{lccccc}
\tablecolumns{6}
\tablewidth{0pc}
\tablecaption{Reduced $\chi^2$ for CXOU J164710.2-455216\tablenotemark{a}
\label{1647_comp_csq}}
\tablehead{
\colhead{Epoch} &
\colhead{$\chi^2_{red}$} &
\colhead{$\chi^2_{red}$} &
\colhead{$\chi^2_{red}$} &
\colhead{$\chi^2_{red}$} &
\colhead{$T$} \\
\noalign{\smallskip}
\colhead{}&
\colhead{(1T)} &
\colhead{(2T)} &
\colhead{(3T)} &
\colhead{(XSPEC)} &
\colhead{(keV)} 
}
\startdata
Sep 06 & 1.05 & 0.86 & 0.31 & 1.24 & - \\
Feb 07 & 1.32 & 0.76 & 0.65 & 0.83 & - \\
Aug 07 & 0.97 & 0.91 & 0.44 & 1.01 & - \\
Feb 08 & 1.45 & 1.12 & 0.63 & 1.08 & $0.62^{+0.06}_{-0.09}$ \\
Aug 08 & 1.45 & 1.23 & 0.79 & 1.23 & - \\
Aug 09 & 2.03 & 1.97 & 1.52 & 1.36 & - \\
\enddata
\tablenotetext{a}{Same as in tab. \ref{1810_comp_csq}}
\end{deluxetable}

\subsubsection{Spectra}\label{1647spectra}

\begin{figure}
\includegraphics[width=2.5in,angle=270]{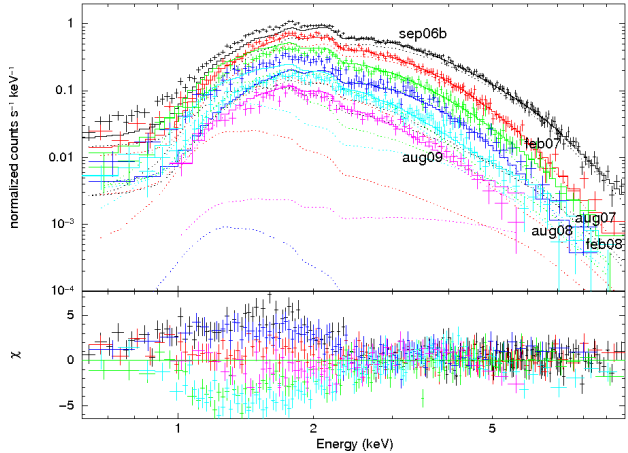}
\caption{Same as in Fig. \ref{1810_spectra} for the six {\it XMM-Newton}
observations of CXOU J164710.2-455216.
\label{1647_spectra}}
\end{figure}

The spectral analysis for CXOU J164710.2-455216 was carried out using
the same approach discussed in \S\ref{1810spec}. We fitted three {\tt ntzang}
components, each representative of an emitting region at different
temperature, and froze all parameters apart from the three
normalizations and $N_H$ (which were forced to be the same for all the
components and for all epochs). Moreover, since the lightcurve
analysis the February 2008 observation failed to provide an
unambiguous value for the hot temperature, also this parameter was
left free to vary. Results are shown in Fig. \ref{1647_spectra}.
Given the approach we used for the fit, the agreement is quite
satisfactory (reduced $\chi^{2}$  are listed if Tab.
\ref{1647_comp_csq}). Systematic residuals at low (1--2 keV)
energies are however present, especially in the September 2006, August 2007 and August 2008
observations. $N_H$ is found to be $(2.14 \pm 0.015)
\times 10^{22} \ {\rm cm^{-2}}$, somewhat higher than that derived
by \cite{naik07}, $N_H = (1.73 \pm 0.03) \times 10^{22} \ {\rm
cm^{-2}}$.

\section{Discussion}\label{discus}

The simultaneous study of the timing and spectral characteristics of the
transient AXPs XTE J1810-197 and CXOU J164710.2-455216 presented in this
paper shows that the post-burst evolution of two sources share a number of
similar properties. In particular, the long-term variability of the pulse
profiles and spectra appears to be (semi)quantitatively consistent with a scenario in which
the star surface thermal distribution and magnetospheric properties progressively change in time.
Our results were derived
within the twisted magnetosphere model for magnetars and support a picture
in which the twist affects only a small bundle of closed field lines around
one of the magnetic poles.
As discussed by \cite{belob09}, if the twist is initially
confined along the magnetic axis, the returning currents hit a limited portion
of the star surface (typically a polar cap), which becomes hotter.
In this scenario the post-outburst evolution is related to
the twist decay, during which the bundle shrinks, and the heated region
decreases both in size and temperature.
We found evidence for a cooling/shrinking of the heated polar cap in both
sources, together with a decrease of the twist angle. It
should be noted that our magnetospheric model assumes a global
twist, since no spectral calculations are currently available for a localized
twist.

Within this common framework, there are nonetheless differences between the
two TAXPs.
For XTE J1810-197 we found that the star thermal map comprises three 
regions:
a hot cap, a surrounding warm corona, and the rest of the surface at
a colder temperature. The hot cap decreases in size and temperature
until it becomes indistinguishable from the corona around March 2006.
Also the warm corona shrinks, although its temperature
stays about constant at $\sim 0.3 \ {\rm keV}$. It becomes a cap in
March 2006 and it is still visible in our last observation
(September 2007) although its size is down to $0.5\%$ of the
entire surface. The rest of the surface remains at a temperature
comparable to the quiescent one (as measured by {\it ROSAT})
during the entire evolution, indicating that the outburst likely
involved only a fraction of the star surface. \cite{bernardini09} obtained
similar results using a 3BB model, although they did not attempt
to locate the different emitting regions on the star surface nor to fit
the pulse profiles. In their (purely spectral) analysis
the hot region is visible slightly longer (until March 2006);
the reason for the difference with respect to our results being most probably
the resolution of our surface grid. Moreover, in our case the hot
temperature decrease is more pronounced.
The twist angle decreases from $\sim 0.8$ rad to $\sim 0.5$ rad during the
first two years, and then it remains roughly constant.

Much as in the case of XTE J1810-197, the thermal map of CXOU 
J164710.2-455216 is well
reproduced by three different regions. However, while the evolution of the hot cap is similar,
i.e. it decreases in size and temperature until it disappears in the August 2008 observation, the behavior
of the warm corona is different. Now the warm temperature remains constant at $\sim 0.45 \ {\rm keV}$ and
the area increases. Actually, the area of the ``hot+warm'' region is constant and covers about $\sim 30\%$
of the surface, while the remaining $\sim 70\%$ is at a constant cooler temperature, $\sim 0.15 \ {\rm keV}$.
This is suggestive of a picture in which the "quiescent'' state of the source is characterized by a two-temperature
map, with a warm polar region superposed to the cooler surface. The outburst might have heated a portion of the
warm cap, producing the hot zone which then cooled off. It is intriguing to notice that the disappearance of the
hot spot occurs at the same time (August 2008) at which the pulse profile dramatically changed, switching from a
three-peaked to a single-peaked pattern. A quasi-sinusoidal shape of the lightcurve was observed when the source
was in quiescence \cite[][]{israel07}. However, at that time the pulsed fraction was nearly 100\% above 4 keV,
likely indicating the presence of a small hot spot which is periodically occulted as the star rotates. This
is in agreement with our finding that this TAXP is a nearly orthogonal rotator. Whether CXOU
J164710.2-455216 is presently approaching quiescence is unclear. If this is the case, its quiescent state is different
from that observed in 2005 and also from that of XTE J1810-197.

It is worth stressing that our claim that the
temperature does not change spatially in each of the regions should not be
taken literally. The assumption that the surface can be divided in three
(or two) thermal regions was mainly introduced to simplify the calculations
while catching the essential features of the model.
A smooth temperature variation within
a zone is likely to be present. However, it is difficult to reconcile the
observed pulsed fraction of XTE J1810-197
in the September 2006 observation ($\gtrsim 10\%$, see Fig. \ref{pf_1810})
even accounting for the temperature gradient
induced by the large-scale dipolar field. This may be an indication that,
as our analysis shows, there is a residual
twist even in the quiescent state.

In this respect we note that our spectral calculation is based on a rather
fine subdivision of the star surface ($50 \times 4$ patches in the final
version of the archive), so we could have produced pulse profiles for
arbitrary complicated thermal maps. The motivation of our choice of the
thermal distribution (a hot polar cap and a warm a corona superimposed to
the colder surface) is threefold: i) a model based on  two thermal
components, originating from a hot cap and a warm corona, was successfully
applied to XTE J1810-197 by \cite{perna08}; ii) inclusion of a third,
colder component in the spectrum of the same source was shown
to be statistically significant by \cite{bernardini09}; and iii)
it is consistent with theoretical predictions for a twisted
magnetosphere in an AXP \cite[][see above]{belob09}. In addition,
this is the simplest map for which we were
able to obtain constant values, to within the errors, for the two
geometrical angles $\chi$ and $\xi$ during
the entire period covered by the observations.

In their analysis of XTE J1810-197, \cite{perna08} assumed that the
X-rays come from two concentric regions with
varying temperatures and areas, each emitting a blackbody spectrum;
the rest of the surface was taken
to be at zero temperature. They derived the angles $\chi$ and $\xi$,
and, although their solution is not unique, they claim that the
pair $\chi \sim 53^{\circ}$, $\xi \sim 23^{\circ}$ is favored.
While this value of $\xi$ coincides with our estimate, the two values of the
inclination of the line-of-sight are
in substantial disagreement. Also the emitting areas of the hot/warm region
and their temperatures turn out to be
different in the two cases. Their estimate of the hot temperature is
always higher than ours and the size of the warm
corona is not monotonically decreasing. We remark that quantitative
differences are to be expected given the different assumed spectral
models (blackbody vs. RCS); moreover
because \cite{perna08} did not include a colder
region\footnote{It was already noted by \cite{bernardini09}
that the addition of the colder component produces a monotonic
decrease in both the hot and warm areas}.

\begin{figure}
\includegraphics[width=3.3in]{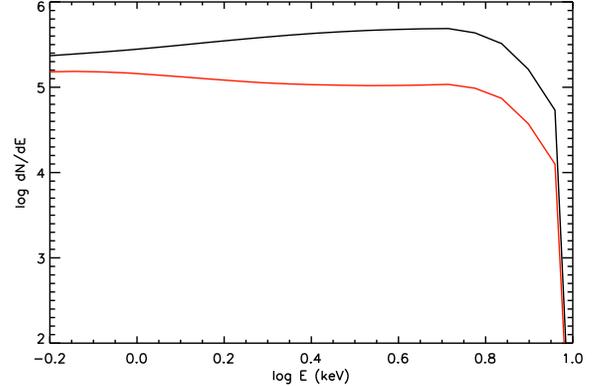}
\caption{Comparison between the spectrum obtained adding three single
{\tt ntzang} model (red) and the spectrum of a neutron star with a
thermal map consisting of three regions at different temperatures (black).
The two spectra are relative to the September 2004 observation of XTE J1810-197.
\label{conf_spectra}}
\end{figure}

Finally, we caveat that our analysis relies on a number of simplifying
assumptions. We already mentioned that the synthetic spectra we used were
obtained with the Monte Carlo code by \cite{nobili08}, which was designed
to solve radiation transport in a globally twisted magnetosphere. Even though
we took thermal photons
to originate mostly in a limited polar region, this does not
self-consistently describe resonant up-scattering in a magnetosphere where only a limited bundle of
field lines is actually twisted, as is probably the case in AXPs \cite[]{belob09}.
Moreover, as we discussed in \S \ref{1810spec}, the {\tt ntzang} XSPEC model is available only
in tabular form and it was created assuming emission at constant temperature from the entire
star surface. As such, it is not suited to be applied directly to the
present case. As a compromise, we decided to fit the spectra
by adding together two/three (absorbed) {\tt ntzang} components,
each associated to one of the emitting regions, at temperatures $T_h$, $T_w$ and $T_c$, respectively.
While this procedure works (and is routinely employed) in the case of blackbody spectra, it is
expected to be only approximately correct when different {\tt ntzang} components are added together. The reason is that
the effects of resonant scattering on thermal photons depends on the location of the  primary emission, since
the magnetospheric electron density is not isotropic.  As a consequence,
assuming thermal emission from a cap of limited size or from the entire
star, even if the two are taken at the same temperature, will give rise to
different spectra. We checked this approximation for all the spectra we analyzed, finding
that the maximum relative error is $\sim 0.6$, while the energy-averaged
error is always between 0.2 and 0.4 both for XTE J1810-197 and CXOU J164710.2-455216.
An example is shown in Fig \ref{conf_spectra}. Although we are
aware that this is
not optimal, it  provides a reasonable way to describe radiation coming from a
magnetar with non-uniform thermal emission within the context of our model.

\section {Conclusions}
\label{conc}

The monitoring of the two TAXPs XTE J1810-197 and CXOU
J164710.2-455216, carried out with {\it XMM-Newton} in recent years,
gave us the possibility to test the twisted magnetosphere
model and understand how the physical parameters in the two
sources change during the post-outburst evolution. We summarize our main findings below,
remarking again that they were obtained under a number of assumptions (e.g. globally twisted field,
three temperature thermal map).

\begin{itemize}
\item{Soon after the outburst onset the surface thermal
distribution in XTE J1810-197 and CXOU J164710.2-455216 is well
described by three components: a hot cap, a surrounding warm
corona while the rest of the neutron star surface is at a lower
temperature.}
\item{The analysis of the pulse profile evolution for
XTE J1810-197 revealed that both the hot cap and the warm corona
decrease in size so that in the last observation (September 2007)
virtually all the neutron star surface emits at a temperature
compatible with the quiescent one.}
\item{The same analysis for CXOU J164710.2-455216 showed
that the hot cap decreases in temperature and size, while the warm
corona remains constant in temperature while it increases in size. In
the last two observations we examined (August 2008 and August
2009) the source thermal map comprises a hot cap covering
$\sim 30 \%$ of the neutron star surface, while the remaining surface is
cooler. There are hints that this could be the
quiescent state of the TAXP.}
\item{For both sources
the twist angle is highest at the outburst
onset and then monotonically decreases in time until it reaches a nearly constant,
non-zero value.}
\item{The same model configuration which best-fits the observed pulse profiles (thermal
map, twist angle, electron bulk velocity, and geometrical angles) provides a reasonable description of
{\it XMM-Newton} spectra in the 0.1--10 keV band for both sources.}
 \end{itemize}

To our knowledge this is the first time that a self-consistent spectral and
timing analysis, based on a realistic modelling of resonant
scattering, was carried out for magnetar sources, considering simultaneously a large
number of datasets over a baseline of years. Present results support to a picture in which
only a limited portion of the magnetosphere was affected by the twist. Future developments will
require detailed spectral calculations in a magnetosphere with a localized twist which decays
in time.

\acknowledgements
We are grateful to an anonymous referee for his/her constructive criticism and helpful suggestions which
helped in improving a previous version of this paper. Work partially supported by INAF-ASI through grant
AAE I/088/06/0.


\begin{thebibliography}{}
%

\bibitem[\protect\citeauthoryear{Beloborodov}{2009}]{belob09}
Beloborodov, A.M. 2009, ApJ, 703, 1044
%
\bibitem[\protect\citeauthoryear{Bernardini et al.}{2009}]{bernardini09}
Bernardini, F., et al. 2009, A\&A, 498, 195
%
\bibitem[\protect\citeauthoryear{Burgay et al.}{2006}]{burgay06}
Burgay, M., Rea, N., Israel, G.L., \& Possenti, A. 2006, ATel, 903
%
\bibitem[\protect\citeauthoryear{Camilo et al.}{2006}]{camilo06}
Camilo, F., Ransom, S.M., Halpern, J.P., Reynolds, J., Helfand, D.J., Zimmerman, N., \&  Sarkissian, J. 2006, Nature, 442, 892
%
\bibitem[\protect\citeauthoryear{Camilo et al.}{2007}]{camilo07a}
Camilo, F., et al. 2007a, ApJ, 669, 561
%
\bibitem[\protect\citeauthoryear{Campana, \& Israel}{2006}]{campana06}
Campana, S., \& Israel G.L. 2006, ATel, 893
%
\bibitem[\protect\citeauthoryear{Duncan, \& Thompson}{1992}]{duncan92}
Duncan R.C., \&  Thompson C. 1992, ApJ, 392, 9
%
\bibitem[\protect\citeauthoryear{Fahlman, \& Gregory}{1981}]{fg81}
Fahlman, G.G., \& Gregory, P.C. 1981, Nature, 293, 202
%
\bibitem[\protect\citeauthoryear{Fernandez, \& Thompson}{2007}]{ft07}
Fernandez R., \& Thompson C. 2007, ApJ, 660, 615
%
\bibitem[\protect\citeauthoryear{Lyutikov, \& Gavriil}{2006}]{lg06}
Lyutikov M., \& Gavriil F.P. 2006, MNRAS, 368, 690
%
\bibitem[\protect\citeauthoryear{Gotthelf et al.}{2004}]{gotthelf04}
Gotthelf, E.V., Halpern, J.P., Buxton, M., \& Bailyn, C. 2004, ApJ, 605, 368
%
\bibitem[\protect\citeauthoryear{Gotthelf, \& Halpern}{2005}]{gotthelf05}
Gotthelf, E.V., \& Halpern, J.P. 2005, ApJ, 632, 1075
%
\bibitem[\protect\citeauthoryear{Gotthelf, \& Halpern}{2007}]{gotthelf07}
Gotthelf, E.V., \& Halpern, J.P. 2007, Ap\&SS, 308, 79
%
\bibitem[\protect\citeauthoryear{Halpern et al.}{2005}]{halpern05}
Halpern, J.P., Gotthelf, E.V., Becker, R.H., Helfand, D.J., \& White, R.L.  2005, ApJ, 632, 29
%
\bibitem[\protect\citeauthoryear{Kaspi et al.}{2003}]{kaspi03}
Kaspi, V.M., Gavriil, F.P., Woods, P.M., Jensen, J.B., Roberts, M.S.E., \& Chakrabarty, D. 2003, ApJ, 588, 93
%
\bibitem[\protect\citeauthoryear{Krimm et al.}{2006}]{krimm06}
Krimm, H., Barthelmy, S., Campana, S., Cummings, J., Israel, G., Palmer, D., \&  Parsons, A. 2006, GCN Circular 5581
%
\bibitem[\protect\citeauthoryear{Ibrahim et al.}{2004}]{ibrahim04}
Ibrahim, A.I., et al. 2004, ApJ, 609, 21
%
\bibitem[\protect\citeauthoryear{Israel et al.}{2004}]{israel04}
Israel, G.L., et al. 2004, ApJ, 603, 97
%
\bibitem[\protect\citeauthoryear{Israel, \& Campana}{2006}]{israel06}
Israel, G.L., \& Campana S. 2006, ATel, 896
%
\bibitem[\protect\citeauthoryear{Israel et al.}{2007}]{israel07}
Israel, G.L., Campana, S., Dall'Osso, S., Muno, M.P., Cummings, J., Perna, R., \&  Stella, L. 2007, ApJ, 664, 448
%
\bibitem[\protect\citeauthoryear{Leahy et al.}{2008}]{le08}
Leahy, D.A., Morsink, S.M., \& Cadeau, C.  2008, ApJ, 672, 1119
%
\bibitem[\protect\citeauthoryear{Leahy et al.}{2009}]{le09}
Leahy, D.A., Morsink, S.M., Chung, Y.-Y., \& Chou, Y. 2009, ApJ,
691, 1235
%
\bibitem[\protect\citeauthoryear{Laros et al.}{1986}]{laros86}
Laros, J.G., Fenimore, E.E., Fikani, M.M., Klebesadel, R.W., \& Barat, C. 1986, Nature, 322, 152
%
\bibitem[\protect\citeauthoryear{Mazets et al.}{1979}]{mazets79}
Mazets, E.P., Golentskii, S.V., Ilinskii, V.N., Aptekar, R.L., \&
Guryan, I.A. 1979, Nature, 282, 587
%
\bibitem[\protect\citeauthoryear{Mereghetti, \& Stella}{1995}]{merste95}
Mereghetti, S., \& Stella, L. 1995, ApJ, 628, 938
%
\bibitem[\protect\citeauthoryear{Mereghetti et al.}{2005}]{mereghetti05b}
Mereghetti, S., et al. 2005, ApJ, 628, 938
%
\bibitem[\protect\citeauthoryear{Mereghetti et al.}{2006}]{mereghetti06}
Mereghetti, S., et al. 2006, A\&A, 450, 759
%
\bibitem[\protect\citeauthoryear{Mereghetti}{2008}]{mereghetti08}
Mereghetti, S. 2008, A\&A Review, 15, 225
%
\bibitem[\protect\citeauthoryear{Mereghetti et al.}{2009}]{mereghetti09}
Mereghetti, S., et al. 2009, ApJ, 696, 74
%
\bibitem[\protect\citeauthoryear{Morrison, \& McCammon}{1983}]{mormc83}
Morrison, R., \& McCammon, D. 1983, ApJ, 270, 119
%
\bibitem[\protect\citeauthoryear{Muno et al.}{2006}]{muno06}
Muno, M.P., et al. 2006, ApJ, 636, 41
%
\bibitem[\protect\citeauthoryear{Muno et al.}{2006b}]{muno06b}
Muno, M.P., Gaensler, B., Clark, J.S., Portegies Zwart, S., Pooley, D., de Grijs, R., Stevens, I., \&  Negueruela, I. 2006, ATel, 902, 1M
%
\bibitem[\protect\citeauthoryear{Muno et al.}{2007}]{muno07}
Muno, M.P., Gaensler, B.M., Clark, J.S., de Grijs, R., Pooley, D., Stevens, I.R., \& Portegies Zwart, S.F. 2007, MNRAS, 378, L44
%
\bibitem[\protect\citeauthoryear{Naik et al.}{2008}]{naik07}
Naik, S., et al, 2008, PASJ, 60, 237
%
\bibitem[\protect\citeauthoryear{Nobili, Turolla, \& Zane}{2008}]{nobili08}
Nobili R., Turolla R., \& Zane S. 2008, MNRAS, 386, 1527
%
\bibitem[\protect\citeauthoryear{Perna \& Gotthelf}{2008}]{perna08}
Perna, R., \& Gotthelf, E.V. 2008, ApJ, 681, 522
%
%\bibitem[\protect\citeauthoryear{Rea et al.}{2004a}]{rea04}
%Rea, N., et al. 2004a, ApJ, 609, 21
%
\bibitem[\protect\citeauthoryear{Rea et al.}{2004}]{rea04b}
Rea, N., et al. 2004b, A\&A 425, 5
%
\bibitem[\protect\citeauthoryear{Rea et al.}{2008}]{rea08}
Rea, N., Zane, S., Turolla, R., \& Lyutikov, M. 2008, ApJ, 686,
1245
%
\bibitem[\protect\citeauthoryear{Testa et al.}{2008}]{testa08}
Testa, V. et al. 2008, A\&A, 482, 607
%
\bibitem[\protect\citeauthoryear{Thompson \& Duncan}{1993}]{thompson93}
Thompson, C., \& Duncan, R.C. 1993, ApJ, 408,194
%
\bibitem[\protect\citeauthoryear{Thompson \& Duncan}{1995}]{td95}
Thompson, C., \& Duncan, R.C. 1995, MNRAS, 275, 255
%
\bibitem[\protect\citeauthoryear{Thompson, Lyutikov, \& Kulkarni}{2002}]{thompson02}
Thompson, C., Lyutikov, M., \& Kulkarni, S.R. 2002, ApJ, 274, 332
%
\bibitem[\protect\citeauthoryear{Woods \& Thompson}{2006}]{woodsthomp06}
Woods, P.M., \& Thompson, C. 2006, in Compact stellar X-ray sources, Lewin, W. and van der Klis, M. Eds.,
Cambridge University Press, Cambridge, UK,  p. 547
%
\bibitem[\protect\citeauthoryear{Zane, \& Turolla}{2006}]{zt06}
Zane, S., \& Turolla, R. 2006, MNRAS, 366, 727
%
\bibitem[\protect\citeauthoryear{Zane et al.}{2009}]{zrtn09}
Zane, S., Rea, N., Turolla, R., \& Nobili, L. 2009, MNRAS, 398, 1403
%
\end{thebibliography}
\end{document}